\documentclass[twocolumn]{aastex62}
\usepackage{hyperref}
\usepackage{amsmath,amstext}
\usepackage[T1]{fontenc}
\usepackage[figure,figure*]{hypcap}
\usepackage{graphicx}
\usepackage{multirow}

\received{June 27, 2019}
\submitjournal{ApJ}

\shorttitle{The Analogous Structure of Black Hole Accretion Flows} 
\shortauthors{Ruan {\it et al.}}
\begin{document}

\title{The Analogous Structure of Accretion Flows in Supermassive and Stellar Mass Black Holes: \\ New Insights from Faded Changing-Look Quasars}

\correspondingauthor{John J. Ruan}
\email{john.ruan@mcgill.ca}

\author[0000-0001-8665-5523]{John~J.~Ruan}
\affil{McGill Space Institute and Department of Physics, McGill University, 3600 rue University, Montreal, Quebec, H3A 2T8, Canada}

\author{Scott~F.~Anderson}
\affil{Department of Astronomy, University of Washington, Box 351580, Seattle, WA 98195, USA}

\author{Michael~Eracleous}
\affil{Department of Astronomy \& Astrophysics and Institute for Gravitation and the Cosmos, The Pennsylvania State University, 525 Davey Lab, University Park, PA 16802, USA}

\author{Paul~J.~Green}
\affil{Harvard Smithsonian Center for Astrophysics, 60 Garden St, Cambridge, MA 02138, USA}

\author{Daryl Haggard}
\affil{McGill Space Institute and Department of Physics, McGill University, 3600 rue University, Montreal, Quebec, H3A 2T8, Canada}
\affil{CIFAR Azrieli Global Scholar, Gravity \& the Extreme Universe Program, Canadian Institute for Advanced Research, 661 University Avenue,
Suite 505, Toronto, ON M5G 1M1, Canada}

\author{Chelsea~L.~MacLeod}
\affil{Harvard Smithsonian Center for Astrophysics, 60 Garden St, Cambridge, MA 02138, USA}

\author{Jessie~C.~Runnoe}
\affil{Department of Astronomy, University of Michigan, 1085 S. University Avenue, Ann Arbor, MI 48109, USA}

\author{Malgosia~A.~Sobolewska}
\affil{Harvard Smithsonian Center for Astrophysics, 60 Garden St, Cambridge, MA 02138, USA}

\begin{abstract}

	Despite their factor of $\sim$10$^8$ difference in black hole mass, several lines of evidence suggest possible similarities between black hole accretion flows in active galactic nuclei (AGN) and Galactic X-ray binaries. However, it is still unclear whether the geometry of the disk-corona system in X-ray binaries directly scale up to AGN, and whether this analogy still holds in different accretion states. We test this AGN/X-ray binary analogy, by comparing the observed correlations between the UV-to-X-ray spectral index ($\alpha_\mathrm{OX}$) and Eddington ratio in AGN to those predicted from observations of X-ray binary outbursts. This approach probes the geometry of their disk-corona systems as they transition between different accretion states. We use new \emph{Chandra} X-ray and ground-based rest-UV observations of faded `changing-look' quasars to extend this comparison to lower Eddington ratios of $<$10$^{-2}$, where observations of X-ray binaries predict a softening of $\alpha_\mathrm{OX}$ in AGN. We find that the observed correlations between $\alpha_\mathrm{OX}$ and Eddington ratio of AGN displays a remarkable similarity to accretion state transitions in prototypical X-ray binary outbursts, including an inversion of this correlation at a critical Eddington ratio of $\sim$10$^{-2}$. Our results suggest that the structures of black hole accretion flows directly scale across a factor of $\sim$10$^8$ in black hole mass and across different accretion states, enabling us to apply theoretical models of X-ray binaries to explain AGN phenomenology.

\end{abstract}

% ============================
\section{Introduction}
\label{sec:intro}

	We observe accreting black holes in two main classes that are distinguished by their vastly different masses. Stellar mass black holes found in X-ray binary systems have typical masses of $\sim$5-15~$M_\odot$ \citep[e.g.,][and references therein]{casares14}, and are observed when the black hole accretes from a companion star. In contrast, supermassive black holes at the centers of galaxies have typical masses of $\sim$10$^{7-9}$~$M_\odot$, and are observed as active galactic nuclei (AGN) when they accrete from their nearby gas-rich environment. Based on observations of both X-ray binaries and AGN, it is now thought that the {\it structure} of accretion flows around both types of black holes depends primarily on the {\it rate} of accretion per unit black hole mass (i.e., Eddington ratio). As the Eddington ratio fluctuates, the accretion flow transitions dramatically into different states, each with distinct geometries and multi-wavelength spectral characteristics.   
		
	Direct observations of accretion state transitions in stellar mass black holes have revealed a rich phenomenology \citep{homan05, remillard06}, that has led to concordant physical models \citep[e.g., see review by ][]{done07}. At high Eddington ratios, the accretion flow in X-ray binaries forms a geometrically-thin accretion disk \citep{shakura73} that emits luminous thermal soft X-rays; this accretion state is known as the high-luminosity/soft-spectrum state. As the Eddington ratio drops, the inner region of the thin disk may progressively evaporate \citep{esin97} into a hot radiatively inefficient accretion flow \citep{shapiro76} that is possibly advection dominated \citep{narayan94}. During this process, the X-ray spectrum becomes dominated by Comptonized hard X-rays (possibly from the advection dominated accretion flow), and the X-ray binary enters a low-luminosity/hard-spectrum state. At a bolometric Eddington ratio below $L_\mathrm{bol}/L_\mathrm{Edd} \lesssim 10^{-2}$ in the low/hard state, X-ray observations become scarce due to the low luminosities. However, in the few outbursts with X-ray observations in this regime, the X-ray spectrum is observed to soften again \citep[e.g.,][]{ebisawa94, revnivtsev00, tomsick01, corbel04, kalemci05, wu08, russell10, homan13, kalemci13, kajava16, plotkin17}, possibly due to the onset of a new dominant emission component \citep{sobolewska11b}, such as cyclo-synchrotron \citep{narayan95, wardzinski00, veledina11} or jet synchrotron \citep{zdziarski03, markoff04, markoff05}. Despite the success of this general picture for accretion state transitions in stellar mass black holes, it remains unclear if supermassive black hole accretion flows undergo similar processes. 	

	Previous observations have revealed evidence that AGN display some characteristics of X-ray binary phenomenology, and interpretations of these similarities often propose that AGN behave like X-ray binaries that are scaled-up in size. For example, the discovery that the radio luminosities, X-ray luminosities, and black hole masses of both low-luminosity AGN and low/hard state X-ray binaries lie on a `fundamental plane' of black hole activity \citep{merloni03, falcke04} suggests some link between weakly-accreting black holes across all mass scales. A second example is the observed relation between characteristic timescales in the X-ray light curves (in the form of breaks in the power spectrum), black hole masses, and X-ray luminosities of luminous AGN and high/soft state X-ray binaries \citep{mchardy06, kording07}, which likely reflects some characteristic size scale in all black hole accretion flows. A third example is the observed correlation between radio loudness and the dominance of accretion disk emission in the spectral energy distributions (SEDs) of both AGN and X-ray binaries, that links the presence of radio jets to the properties of the accretion flow in all accreting black holes \citep{kording06}. Finally, several authors have pointed out similarities in the correlations between the X-ray photon index with Eddington ratio in both X-ray binaries and AGN \citep{yang15}, which also suggests some link between the Comptonized hard X-ray emission in AGN and X-ray binaries. Despite these various lines of evidence, many open questions still persist over whether black hole accretion flows are indeed scale-invariant.
	
	A key uncertainty in the AGN/X-binary analogy is whether this analogy holds across all accretion states and Eddington ratios. In other words, it is still unclear if the geometry of the disk-corona system in AGN undergoes similar evolution as a function of their Eddington ratio during state transitions, in comparison to X-ray binary outbursts. Observationally confirming this fundamental property that underpins the AGN/X-ray analogy would be powerful, as it enables us to apply our understanding of X-ray binaries to explain AGN phenomenology (and vice versa). However, direct comparisons between the accretion flow geometry in AGN and X-ray binaries during outbursts have thus far been difficult, since detailed observations of accretion state transitions in individual AGN are scarce. This may be because most AGN accretion state transitions occur over timescales that are too long for direct observations; a simple linear scaling with black hole mass suggests that the $\sim$few days timescales for state transition in outbursting X-ray binaries would occur in AGN over timescales of $\sim$10$^{4-5}$~years. Thus, direct evidence for such an analogy in the underlying geometry of black hole accretion flows has remained observationally elusive.

	To test whether the geometry of accretion flows undergo analogous state transitions in both stellar mass and supermassive black holes, we compare multi-epoch observations of a prototypical X-ray binary outburst to single-epoch observations of a {\it sample} of AGN that spans a wide range of Eddington ratios. In X-ray binaries, the thermal emission from the thin disk peaks in the soft X-rays, while the Comptonized coronal emission dominates the hard X-rays. Thus, the evolving geometry of the disk-corona system during state transitions may be probed using the X-ray spectral index $\Gamma$. In contrast, thin disks in AGN are cooler and so their thermal emission peaks in the UV, while the Comptonized emission dominates the X-rays. Thus, we can similarly probe the structure of the accretion flow using the UV-to-X-ray spectral index ($\alpha_\mathrm{OX}$) between 2500~\AA~and 2~keV. To directly compare the accretion flow geometries of X-ray binaries to AGN, we can map the observed evolution of $\Gamma$ as a function of $L_\mathrm{bol}/L_\mathrm{Edd}$ in individual X-ray binary outbursts to a predicted correlation between $\alpha_\mathrm{OX}$ and $L_\mathrm{bol}/L_\mathrm{Edd}$ in single-epoch observations of a large sample of AGN \citep{sobolewska11a}. Thus, our approach here is to measure the correlation between $\alpha_\mathrm{OX}$ and $L_\mathrm{bol}/L_\mathrm{Edd}$ in a sample of AGN with a wide range of Eddington ratios, and compare this observed correlation to the spectral evolution of a X-ray binary undergoing accretion state transitions during a prototypical outburst.

	Previous observations of AGN that investigate correlations between $\alpha_\mathrm{OX}$ and Eddington ratio have revealed some similarities with X-ray binary outbursts at high $L_\mathrm{bol}/L_\mathrm{Edd}$, but these comparisons have not been possible below the critical $L_\mathrm{bol}/L_\mathrm{Edd}$ $\lesssim$ $10^{-2}$ where an inversion in this correlation is predicted to occur. At higher Eddington ratios of $L_\mathrm{bol}/L_\mathrm{Edd}\gtrsim10^{-2}$, single-epoch X-ray and UV observations of large samples of AGN have previously revealed a hardening of $\alpha_\mathrm{OX}$ as $L_\mathrm{bol}/L_\mathrm{Edd}$ drops from $\sim$1 to $\sim$$10^{-2}$ \citep[e.g.,][]{vignali03, strateva05, steffen06, just07, grupe10, jin12, wu12, vagnetti13, trichas13}. This correlation was also observed in multi-epoch UV/X-ray observations of the fading of Mrk 1018 \citep{noda18}, which confirms this behavior in an individual AGN. However, the predicted softening of $\alpha_\mathrm{OX}$ below $L_\mathrm{bol}/L_\mathrm{Edd}$ $\lesssim$$10^{-2}$ (thus causing an inversion in the correlation between $\alpha_\mathrm{OX}$ and $L_\mathrm{bol}/L_\mathrm{Edd}$) has not been previously observed. This is primarily due to the difficulty of robustly measuring both $\alpha_\mathrm{OX}$ and $L_\mathrm{bol}/L_\mathrm{Edd}$ for AGN below $L_\mathrm{bol}/L_\mathrm{Edd}\lesssim10^{-2}$, for three main reasons. First, at low Eddington ratios, AGN are often dust obscured \citep{fabian08}, and thus measuring their intrinsic UV luminosities (and $\alpha_\mathrm{OX}$) is difficult. Second, broad emission lines often disappear in low luminosity AGN below $L_\mathrm{bol}/L_\mathrm{Edd}\lesssim10^{-2}$, making it difficult to measure $M_\mathrm{BH}$ (and $L_\mathrm{Edd}$). Third, using a sample of AGN with a wide range of Eddington ratios to trace how $\alpha_\mathrm{OX}$ changes as a function of $L_\mathrm{bol}/L_\mathrm{Edd}$ can be hampered by the $T \propto M_\mathrm{BH}^{-1/4}$ scaling of the thin disk temperature with $M_\mathrm{BH}$ at a fixed Eddington ratio. If the AGN sample has a large range in $M_\mathrm{BH}$, this can cause an additional scatter in $\alpha_\mathrm{OX}$. Thus, ideally we would use a sample of AGN with a narrow range in $M_\mathrm{BH}$, but the difficulty of measuring $M_\mathrm{BH}$ at $L_\mathrm{bol}/L_\mathrm{Edd}\lesssim10^{-2}$ also hampers the construction of such a sample. In this paper, we will use a new method to bypass all these issues, with the goal of extending this spectral comparison between X-ray binaries and AGN to $L_\mathrm{bol}/L_\mathrm{Edd}\lesssim10^{-2}$.
		
\begin{figure*} [t!]
\center{
\includegraphics*[width=0.45\textwidth]{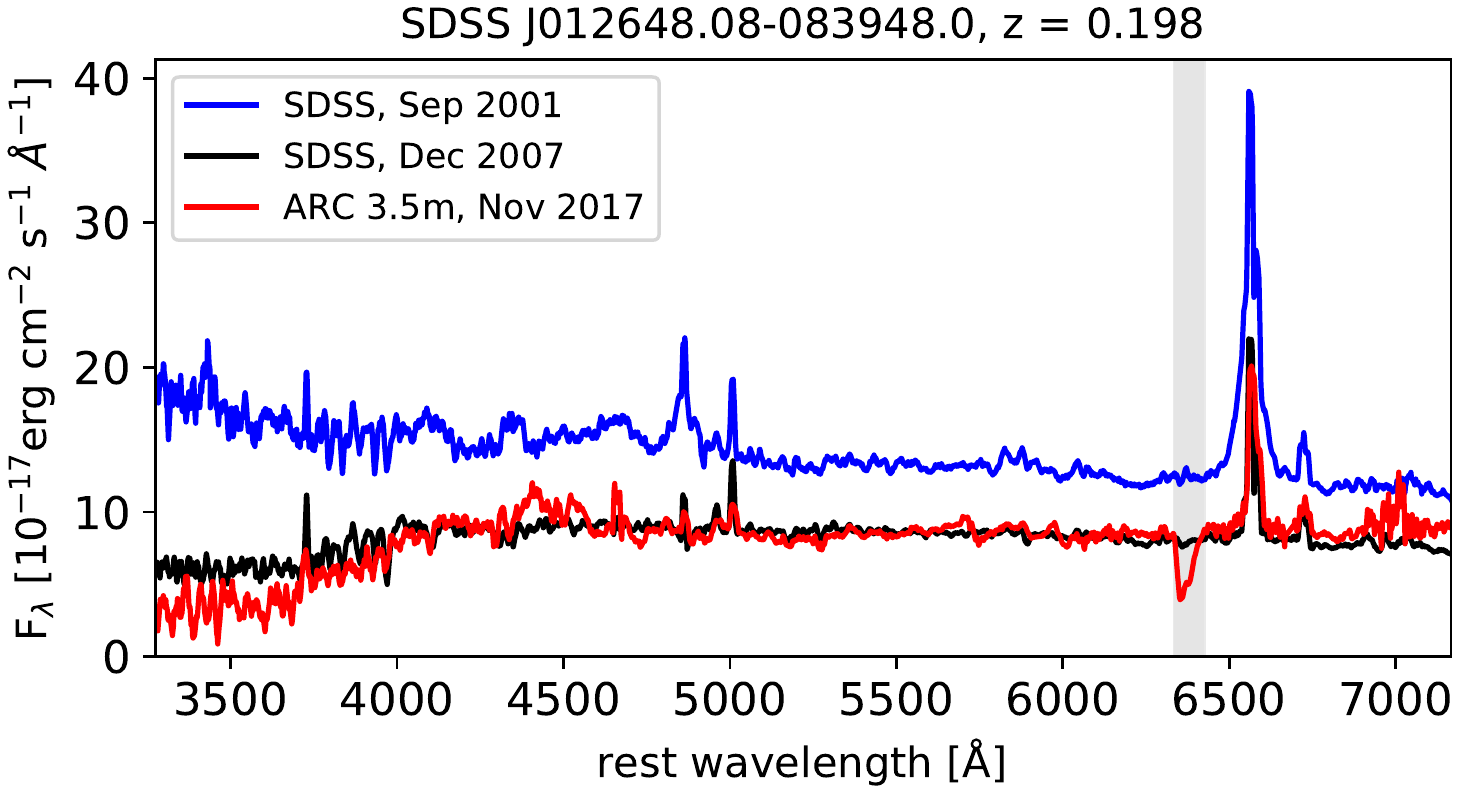} \hspace{5pt}
\includegraphics*[width=0.45\textwidth]{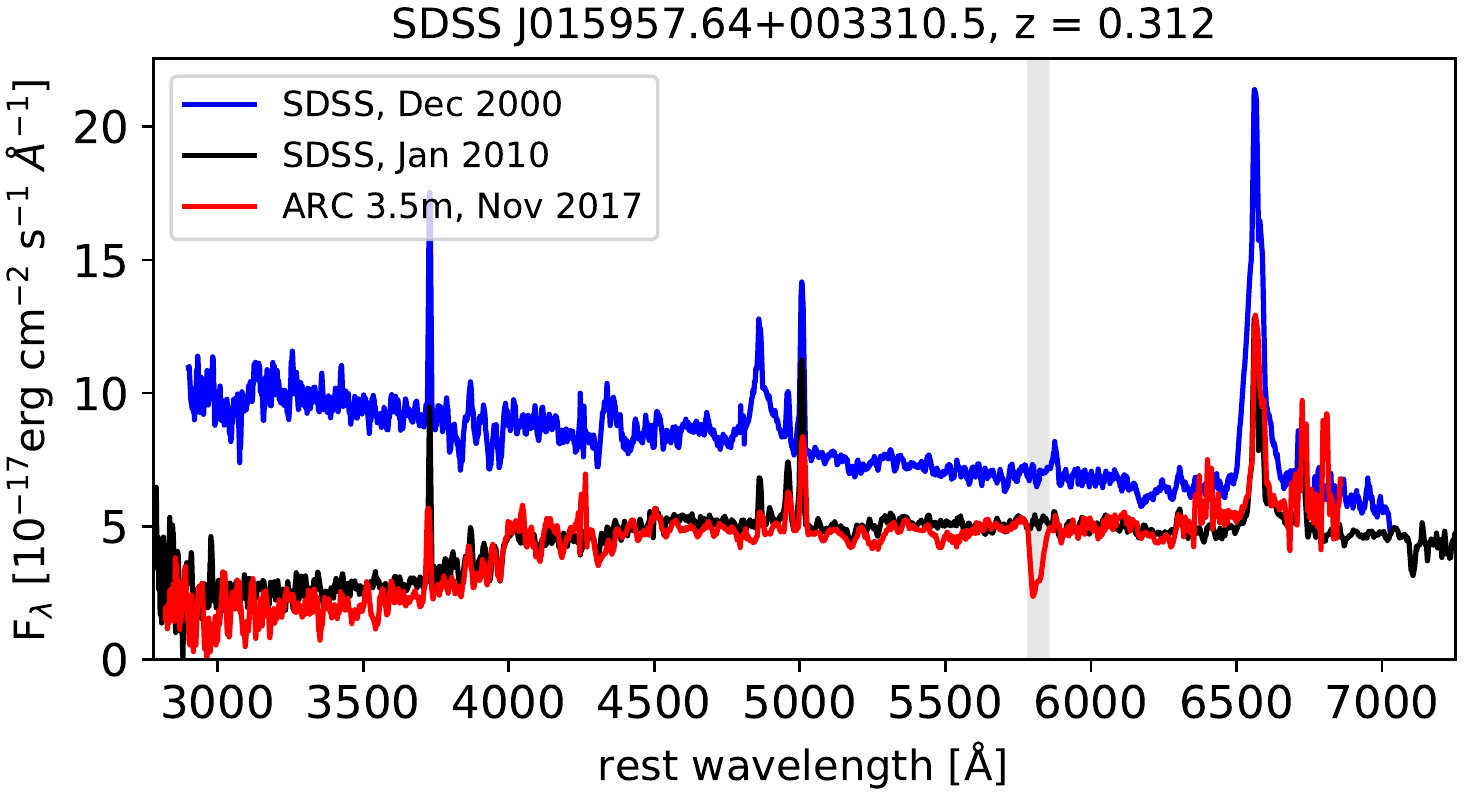}\\ \vspace{5pt}
\includegraphics*[width=0.45\textwidth]{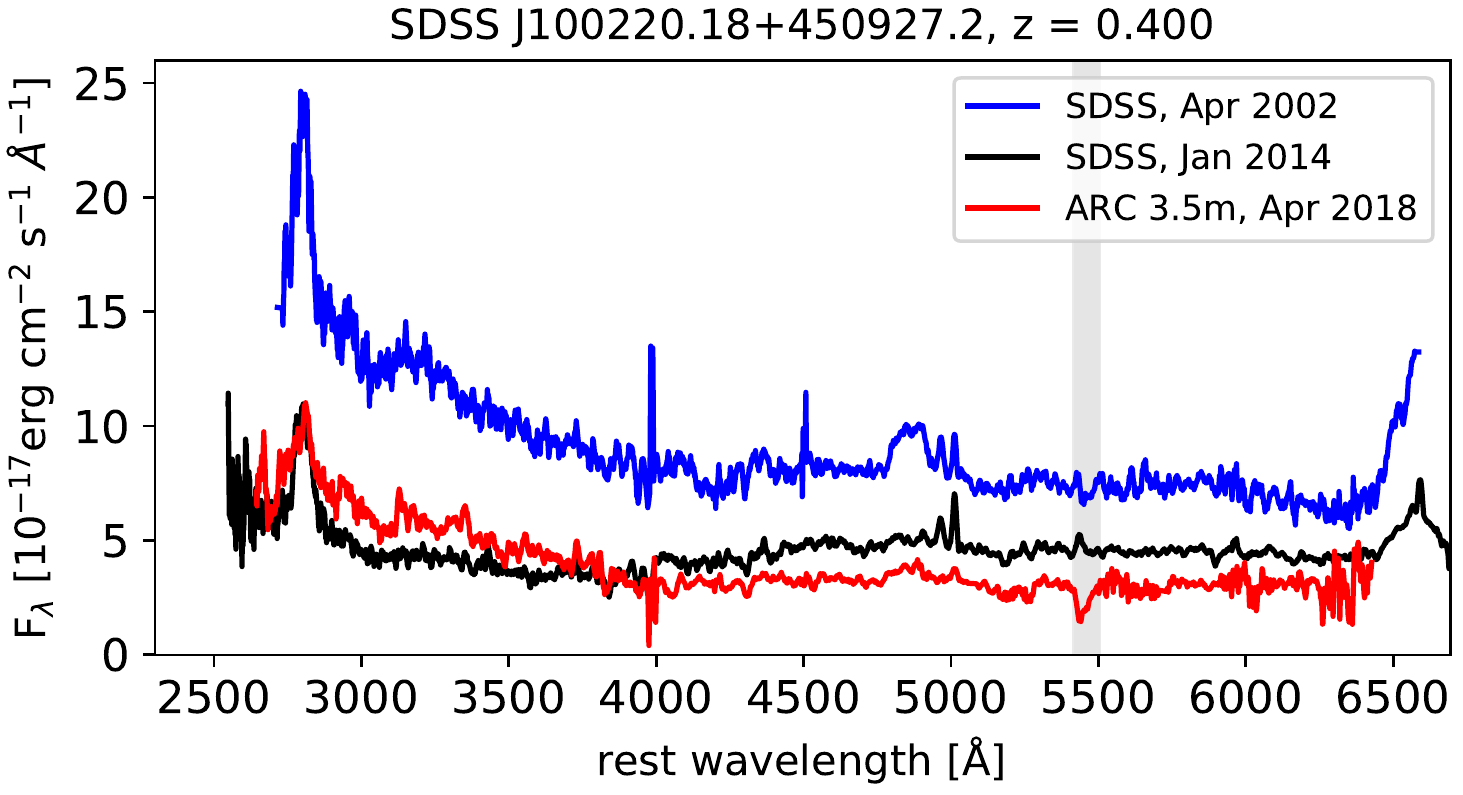} \hspace{5pt}
\includegraphics*[width=0.45\textwidth]{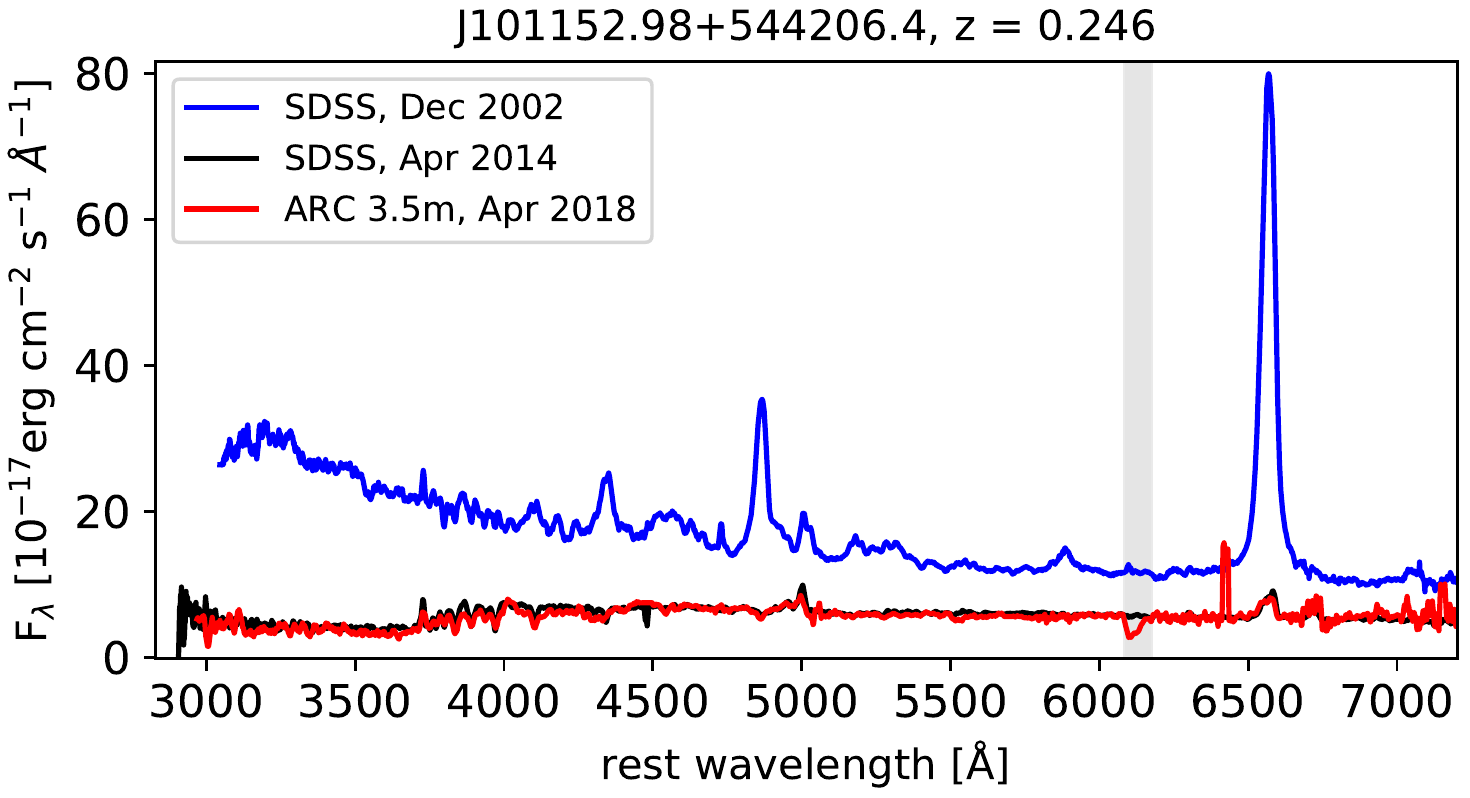}\\ \vspace{5pt}
\includegraphics*[width=0.45\textwidth]{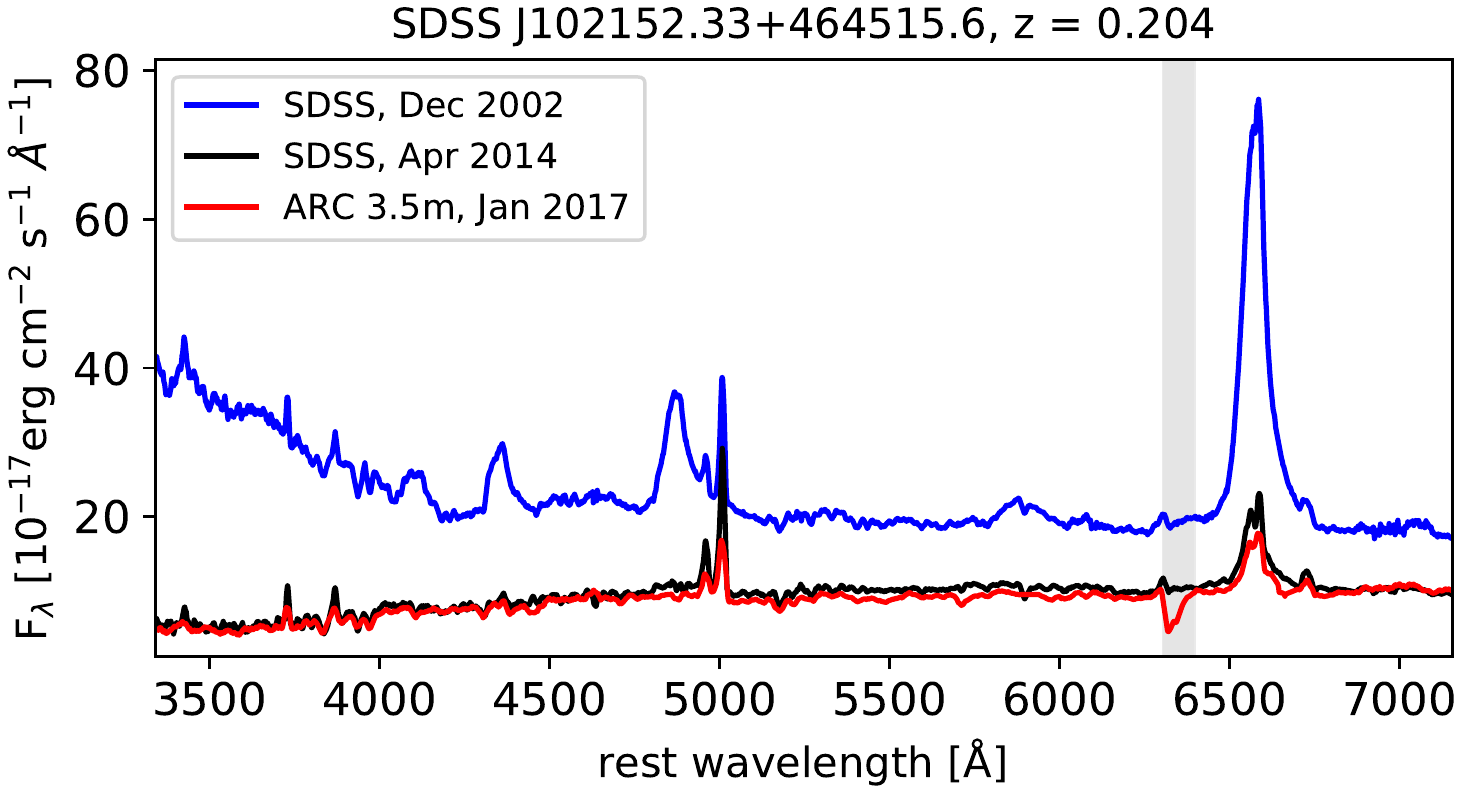} \hspace{5pt}
\includegraphics*[width=0.45\textwidth]{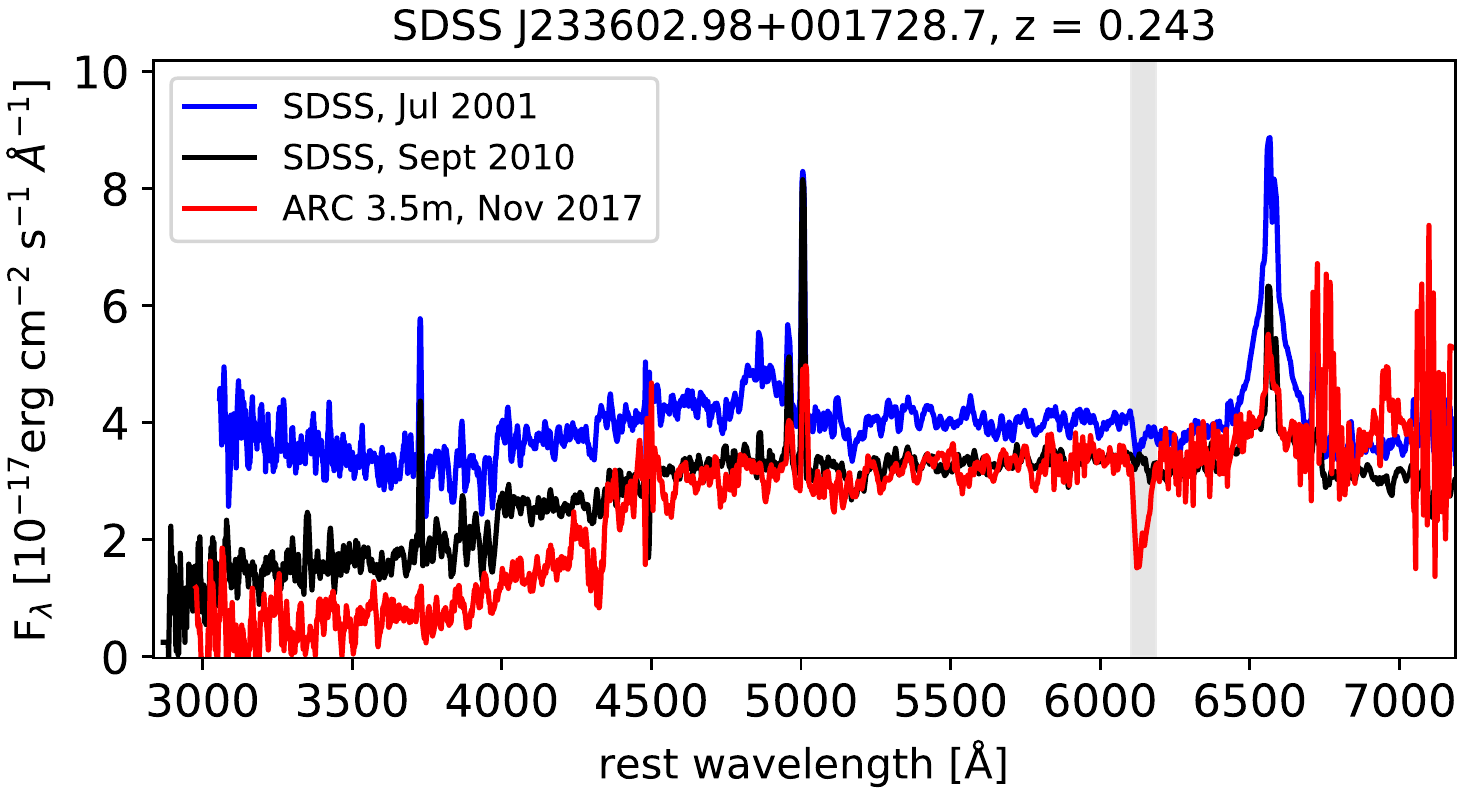}
}
\figcaption{{\bf Multi-epoch optical spectroscopy reveals the dramatic fading of changing-look quasars.} 
Repeat SDSS spectra of six changing-look quasars over $\sim$10 years shows a fading of the broad emission lines and continuum from a bright state (blue) to a faint state (black), consistent with a decrease in their Eddington ratios. A more recent spectroscopic epoch from the ARC 3.5m telescope is also shown (red), which verifies that these changing-look quasars are still in a faint state contemporaneous with our  {\it Chandra} X-ray observations. Wavelength regions affected by telluric absorption are shaded grey.}
\label{fig:1}
\end{figure*}

	The key to our approach is to include faded `changing-look' quasars in our AGN sample, which allows us to extend measurements of $\alpha_\mathrm{OX}$ to low Eddington ratios ($L_\mathrm{bol}/L_\mathrm{Edd} \lesssim 10^{-2}$). Faded changing-look quasars are a class of AGN that are characterized by dramatic drops in the luminosities of their broad emission lines and continuum in repeat optical spectroscopy (see Figure~\ref{fig:1}) \citep[e.g.,][]{lamassa15, ruan16, runnoe16, macleod16, yang18, macleod19, graham19a, sheng19, frederick19}. We emphasize that although we use observations of changing-look quasars in both their bright and faint states, we are {\it not} directly probing the evolution of their $\alpha_\mathrm{OX}$ as they fade, since X-ray detections are not available for the majority of our changing-look quasars in their former bright state. Instead, we investigate the correlation between $\alpha_\mathrm{OX}$ and $L_\mathrm{bol}/L_\mathrm{Edd}$ of our changing-look quasars in their faint state, which can reveal whether the predicted inversion in this correlation is observed at low Eddington ratios of $L_\mathrm{bol}/L_\mathrm{Edd}\lesssim10^{-2}$ that were inaccessible in previous studies. Crucially, our use of changing-look quasars enables us to bypass both the issues of dust obscuration and $M_\mathrm{BH}$ measurements. Since changing-look quasars are observed to be Type 1 AGN before their fading, and their fading has been shown to be consistent with a decrease in their Eddington ratio while disfavoring dust obscuration \citep{lamassa15, ruan16, runnoe16, macleod16, hutsemekers17, sheng17, yang18, stern18, ross18, macleod19, hutsemekers19}, we know that their UV luminosities after they fade are also unobscured. Furthermore, changing-look quasars also allow us to estimate $M_\mathrm{BH}$ for AGN at $L_\mathrm{bol}/L_\mathrm{Edd} \lesssim 10^{-2}$, since we can estimate $M_\mathrm{BH}$ using the prominent broad emission lines in their bright state optical spectra, and then measure both their $\alpha_\mathrm{OX}$ and $L_\mathrm{bol}/L_\mathrm{Edd}$ after they fade.
	
	The outline of this paper is as follows: In Section~2, we describe the new and archival data of our sample of six faded changing-look quasars (including X-ray observations and optical spectroscopy), and our reduction of these data. In Section~3, we describe our modeling of the SDSS optical spectra. In Section~4, we calculate the key parameters of interest for each of our changing-look quasars, including $\alpha_\mathrm{OX}$ and $L_\mathrm{bol}/L_\mathrm{Edd}$ values. In Section~5, we present the observed correlation between $\alpha_\mathrm{OX}$ and $L_\mathrm{bol}/L_\mathrm{Edd}$, and compare to predictions from X-ray binary outbursts. We briefly summarize and conclude in Section~6. Throughout this work, we assume a standard $\Lambda$CDM cosmology with $\Omega_\mathrm{m} = 0.309$, $\Omega_\Lambda = 0.691$, and $H_0 = 67.7$ km s$^{-1}$ Mpc$^{-1}$ \citep{bennett14}.
	
%------- FIGURE 1 -------
% ============================

%%%%%%%%%%%%%%%%%%%%%%%%%%%%%%%%%%%%%%
%%%%%%%%%%%%%%%%%%%%%%%%%%%%%%%%%%%%%%

\section{Data and Reduction}
\label{sec:data}

	For AGN at high Eddington ratios, we use a sample of 150 broad-line AGN from the XMM-COSMOS survey \citep{cappelluti09}, each with spectroscopic $M_\mathrm{BH}$ estimates, as well as measurements of their $\alpha_\mathrm{OX}$ and $L_\mathrm{bol}/L_\mathrm{Edd}$ \citep{lusso10}. For AGN at low Eddington ratios, we use a sample of six changing-look quasars (listed in Table~1) that were previously discovered to have undergone dramatic fading from a bright state to a current faint state in repeat optical spectroscopy \citep[shown in Figure~\ref{fig:1};][]{ruan16, runnoe16, macleod16} from the Sloan Digital Sky Survey \citep[SDSS;][]{york00}. Critically, the XMM-COSMOS AGN and changing-look quasar samples have narrow $M_\mathrm{BH}$ distributions that are nearly identical. 
		
	For the X-ray luminosities needed to 	calculate the $\alpha_\mathrm{OX}$ of our changing-look quasars in their current faint state, we first obtain {\it Chandra X-ray Observatory} observations. We describe these {\it Chandra} observations and our reduction of the X-ray data in Section~2.1. To verify that each changing-look quasar remains in a faint state during these  {\it Chandra} observations, we obtain an additional, contemporaneous optical spectrum from the Astrophysical Research Consortium (ARC) 3.5m telescope. These ARC~3.5m spectra are also shown in Figure~\ref{fig:1}, and we describe these observations and our reduction of the optical spectra in Section~2.2. 
	
	Since only one of our changing-look quasars (J1059) has an archival X-ray detection in its former bright state, we are unable to also measure the bright state $\alpha_\mathrm{OX}$ for most of our changing-look quasars. Instead, we rely on the XMM-COSMOS AGN sample to probe the correlation between $\alpha_\mathrm{OX}$ and $L_\mathrm{bol}/L_\mathrm{Edd}$ in more luminous AGN. Nevertheless, we derive lower limits on the bright state $\alpha_\mathrm{OX}$ using archival X-ray observations from {\it XMM-Newton} and the {\it ROSAT} All-Sky Survey. We describe these bright state X-ray observations in Section~2.3. 
	
%%%%%%%%%%%%%%%%%%%%%%%%%%%%%%%%%%%%%%

\subsection{Faint state X-ray fluxes from  {\it Chandra}}
\label{ssc:chandra}

	To measure the X-ray fluxes of our changing-look quasars after they have faded to a faint state, we obtain new  {\it Chandra} X-ray observations. The details of these observations are listed in Table~2, and their derived faint state luminosities ($\nu L_\mathrm{2keV}$) that are used to calculate $\alpha_\mathrm{OX}$ values are listed in Table~5. 
		
	For five of our sample of six changing-look quasars, we obtain new  {\it Chandra} X-ray observations through a  {\it Chandra} Cycle 18 Guest Observer program (PI: Ruan, Program Number: 18700505). For the one other changing-look quasar in our sample (J0159), we use similar  {\it Chandra} observations that were obtained in a separate Cycle 17 GTO program (PI: Predehl, Program Number: 17700782). All observations are obtained using the ACIS-S3 chip in VFAINT mode, and their exposure times are listed in Table~2. We use the \texttt{CIAO} v.4.9 (CALDB v4.7.7) software \citep{fruscione06} for reduction and analysis of the resultant X-ray data.

	We reprocess the level 2 events files and use \texttt{CIAO}'s {\it repro} script to apply the latest calibrations. We then generate both a 0.5 - 7 keV X-ray counts image and a PSF image of the ACIS-S3 chip. To perform source detection and obtain X-ray positions, we use the  {\it wavedetect} script. All of our six sources are detected in the   {\it Chandra} observations, and each of their measured X-ray positions are within $0.\!\!^{\prime\prime}6$ of their optical positions from SDSS imaging. Since X-ray emission from our changing-look quasars is expected to be point-like, we use the  {\it srcflux} tool to extract source counts at the X-ray position of each object. The source extraction region radii are set to encompass 90\% of the PSF at 1 keV, and the PSF contributions in the source and background regions are estimated using the  {\it arfcorr} tool. The resultant 0.5 - 7 keV source count rates for the changing-look quasars are listed in Table~2.

	For five of our six objects, the source counts are insufficient to extract a high-quality X-ray spectrum, and thus we assume a fixed spectral model to compute fluxes. Specifically, we assume a power-law spectral model with $\Gamma = 1.8$, typical for low-luminosity AGN \citep{gu09, constantin09, younes11}. We test the effects of adopting a range in  $\Gamma$ from 1.6 to 2.0, and find that this produced systematic uncertainties on the resulting X-ray fluxes of $\lesssim$5$\%$ (and $\ll$1$\%$ on the resulting $\alpha_\mathrm{OX}$). Thus, the uncertainties on our X-ray fluxes are dominated by measurement uncertainties, and we ignore the negligible systematics that stem from our assumptions for $\Gamma$. In our spectral model, we include Galactic absorption based on neutral hydrogen column densities from \citet{dickey90}. For one object (J1002), the source counts are sufficient for us to extract and fit its X-ray spectrum. We use the \texttt{Sherpa} software \citep{freeman01, doe07} to fit a power-law spectral model with Galactic absorption, using atomic cross sections from \citet{verner96} and abundances from \citet{wilms00}. The best-fitting spectral model has  $\Gamma = 1.8 \pm 0.1$, and we compute the X-ray flux for J1002 from this fit. The resultant 0.5 - 7 keV unabsorbed model fluxes for our all changing-look quasars are listed in Table~2. Finally, we calculated X-ray luminosities at rest-frame 2~keV using the WebPIMMs\footnote{http://cxc.harvard.edu/toolkit/pimms.jsp} tool, and these $\nu L_\mathrm{2keV}$ measurements are listed in Table~5. 
	
%%%%%%%%%%%%%%%%%%%%%%%%%%%%%%%%%%%%%%

%%%  TABLE: 1 %%%%%%%%%%%%%%%%%%%%%%%%%%%%%%%%%%%%%%

\begin{deluxetable*}{ccccccc}
\centering
\tablecaption{{\bf SDSS spectroscopic properties of changing-look quasars in both their bright and faint states.} Columns include the object name, redshift, observation date of each SDSS spectrum, the optical luminosity state of the changing-look quasar revealed by the SDSS spectrum, the measured broad emission line FWHM (from H$\beta$ or H$\alpha$), and the derived black hole mass. All uncertainties are at 1$\sigma$ confidence level.
}
\tablehead{  
\colhead{Object} & \colhead{$z$} & \colhead{Observation} & \colhead{Luminosity} & \colhead{log($\lambda L_\mathrm{5100\text{\normalfont\AA}}$)}  & \colhead{Broad Line} & \colhead{$M_\mathrm{BH}$}\\ 
\colhead{(SDSS)} & & \colhead{Date} & \colhead{State} & \colhead{[erg s$^{-1}$]} & \colhead{FWHM} & \colhead{[10$^{8}$ $M_\odot$]} \\
& & \colhead{[MJD]} & &  & \colhead{[km s$^{-1}$]} &
 }
\startdata
 \vspace{2pt}
J0126$-$0839    &   0.198   &   52163    &   bright    &  43.7$\pm$0.1  &  4100$\pm$300$^\mathrm{a}$ &  1.2$\pm$0.2 \\
 \vspace{2pt}
 & & 54465 & faint & 42.1$\pm$0.1 & & \\
 \vspace{2pt}
J0159+0033   &   0.312   &   51871    &   bright     &   43.9$\pm$0.1    & 3800$\pm$200$^\mathrm{a}$ &  1.4$\pm$0.2 \\
 \vspace{2pt}
  & & 54465 & faint & 42.7$\pm$0.1 & & \\
 \vspace{2pt}
J1002+4509   &   0.400   &   52376    &   bright      &   44.1$\pm$0.1    &  7300$\pm$1100$^\mathrm{b}$ &  5.0$\pm$1.5 \\
 \vspace{2pt}
  & & 54465 & faint & 43.8$\pm$0.1 & & \\
 \vspace{2pt}
J1011+5442   &   0.246   &   52652     &   bright     &   43.9$\pm$0.1  &   5200$\pm$600$^\mathrm{a}$ &  2.5$\pm$0.7 \\
 \vspace{2pt}
  & & 54465 & faint & 42.6$\pm$0.1 & & \\
 \vspace{2pt}
J1021+4645   &   0.204   &   52614    &    bright     &   43.9$\pm$0.1    &  4800$\pm$300$^\mathrm{a}$ &  2.1$\pm$0.3 \\
 \vspace{2pt}
  & & 54465 & faint & 43.2$\pm$0.1 & & \\
 \vspace{2pt}
J2336+0017   &   0.243   &   52096    &    bright     &    43.3$\pm$0.1    &   6300$\pm$800$^\mathrm{a}$ &  1.8$\pm$0.5 \\
 \vspace{2pt}
  & & 54465 & faint & 42.8$\pm$0.1 & &
\enddata
\tablenotetext{a}{Broad H$\alpha$ emission line} 
\tablenotetext{b}{Broad H$\beta$ emission line} 
\label{tab:tab1}
\end{deluxetable*}

%%%  TABLE: 2 %%%%%%%%%%%%%%%%%%%%%%%%%%%%%%%%%%%%%%
\begin{deluxetable*}{cccccc}
\centering
\tablecaption{{\bf  {\it Chandra} X-ray properties of changing-look quasars in their current faint state.} Columns include the object name, observation date, the  {\it Chandra} ObsID of the exposure, exposure time, count rate, and unabsorbed model flux. All uncertainties are at 1$\sigma$ confidence level.
}
\tablehead{  
\colhead{Object} & \colhead{Observation} & \colhead{Chandra} & \colhead{Exposure} & \colhead{Count Rate}  & \colhead{Unabsorbed Flux} \\ 
\colhead{(SDSS)} & \colhead{Date} & \colhead{ObsID} & \colhead{Time} & \colhead{(0.5 - 7 keV)} & \colhead{(0.5 - 7 keV)} \\
 & \colhead{[MJD]} & & \colhead{[ks]} & \colhead{[10$^{-3}$ cts s$^{-1}$]} & \colhead{[10$^{-14}$ erg s$^{-1}$ cm$^{-2}$]}
 }
\startdata
 \vspace{2pt}
J0126$-$0839    &   57978   &   19516    &   6.9     &   1.8$^{+0.6}_{-0.5}$   &  2.0$^{+0.7}_{-0.5}$  \\
 \vspace{2pt}
J0159+0033   &   57640   &   18639    &   20.7   &   3.2$^{+0.4}_{-0.4}$    &  3.2$^{+0.4}_{-0.4}$ \\
 \vspace{2pt}
J1002+4509   &   58131   &   19515    &   32.6   &   15.0$^{+0.8}_{-0.7}$   &   16.6$^{+0.8}_{-0.8}$ \\
 \vspace{2pt}
J1011+5442   &   58180   &   19518    &   10.9   &    0.5$^{+0.3}_{-0.2}$   &   0.6$^{+0.3}_{-0.2}$ \\
 \vspace{2pt}
J1021+4645   &   58049   &   19514    &   6.9     &    26.5$^{+0.2}_{-0.2}$    &  28.6$^{+2.3}_{-2.2}$ \\
 \vspace{2pt}
J2336+0017   &   57958   &   19517    &   10.9   &    0.5$^{+0.3}_{-0.2}$    &   0.6$^{+0.3}_{-0.2}$ \\
\enddata
\label{tab:tab2}
\end{deluxetable*}

%%%  TABLE: 3 %%%%%%%%%%%%%%%%%%%%%%%%%%%%%%%%%%%%%%
\begin{deluxetable*}{cccccc}
\centering
\tablecaption{{\bf APO spectroscopic observations of changing-look quasars.} Columns include the object name, observation date of the APO spectrum, exposure time, airmass, seeing, and the star used for spectrophotometric calibration.
}
\tablehead{  
\colhead{Object} & \colhead{Observation Date} & \colhead{Exposure Time} & \colhead{Airmass} & \colhead{Seeing}  & \colhead{Calibration Star}\\ 
\colhead{(SDSS)} & \colhead{[MJD]} & \colhead{[s]} &  & \colhead{[arcsec]} &
 }
\startdata
 \vspace{2pt}
J0126$-$0839    &   58084   &   5$\times$600    &   1.35     &  1.2   &  BD+28 4211  \\
 \vspace{2pt}
J0159+0033   &   58084   &   5$\times$600    &   1.25    &   1.2    &  BD+28 4211 \\
 \vspace{2pt}
J1002+4509   &   58217   &   5$\times$900    &   1.03    &   1.6    &  Feige 34 \\
 \vspace{2pt}
J1011+5442   &   58217   &   5$\times$900     &   1.09    &   1.6   &   Feige 34 \\
 \vspace{2pt}
J1021+4645   &   57781   &   3$\times$900    &    1.12    &   1.5    &  Feige 34 \\
 \vspace{2pt}
J2336+0017   &   58079   &   5$\times$600    &    1.27   &    1.2    &   BD+28 4211 \\
\enddata
\label{tab:tab3}
\end{deluxetable*}

%%%  TABLE: 4 %%%%%%%%%%%%%%%%%%%%%%%%%%%%%%%%%%%%%%
\begin{deluxetable*}{cccccc}
\centering
\tablecaption{{\bf  {\it ROSAT} and  {\it XMM-Newton} X-ray properties of changing-look quasars in their former bright state.} Columns include the object name, observation date of the X-ray observation, the telescope used for the observation, exposure time, count rate, and unabsorbed model flux. All uncertainties are at 1$\sigma$ confidence level, and upper limits are 3$\sigma$.
}
\tablehead{  
\colhead{Object} & \colhead{Observation} & \colhead{Telescope} & \colhead{Exposure} &  \colhead{X-ray}  & \colhead{Unabsorbed} \\ 
\colhead{(SDSS)} & \colhead{Date} & & \colhead{Time} & \colhead{Count Rate} & \colhead{X-ray Flux} \\
& \colhead{[MJD]} & & \colhead{[ks]} & \colhead{[10$^{-1}$ cts s$^{-1}$]} & \colhead{[10$^{-14}$ erg s$^{-1}$ cm$^{-2}$]}
 }
\startdata
\vspace{2pt} 
J0126$-$0839    &   48083   &   ROSAT  & 0.4 &   $<$3.5$^\mathrm{a}$  &    $<$7.5$^\mathrm{a}$ \\
\vspace{2pt}
J0159+0033   &   51732   &   XMM  & 10 &   0.2$^\mathrm{b}$   &   26.4$^{+1.4}_{-1.6}$$^\mathrm{b}$ \\
\vspace{2pt}
J1002+4509   &   48191   &   ROSAT   & 0.5  &  $<$ 4.0$^\mathrm{a}$  &   $<$1.4$^\mathrm{a}$ \\
\vspace{2pt}
J1011+5442   &   48177   &   ROSAT   & 0.5  &  $<$ 1.1$^\mathrm{a}$   &   $<$0.45$^\mathrm{a}$ \\
\vspace{2pt}
J1021+4645   &   48167   &   ROSAT   & 0.5  &  $<$ 4.2$^\mathrm{a}$  &   $<$4.7$^\mathrm{a}$ \\
\vspace{2pt}
J2336+0017   &   48224   &   ROSAT   & 0.4  &  $<$ 1.6$^\mathrm{a}$   &   $<$3.5$^\mathrm{a}$ \\
\enddata
\tablenotetext{a}{0.1 - 2.4 keV 3$\sigma$ limit} 
\tablenotetext{b}{2 - 10 keV} 
\label{tab:tab4}
\end{deluxetable*}

%%%%%%%%%%%%%%%%%%%%%%%%%%%%%%%%%%%%%%

\subsection{ARC 3.5m optical spectra}
\label{ssc:APO}

	To verify that our faded changing-look quasars are still in a faint state during the {\it Chandra} observations and have not re-brightened in the optical, we obtain a more recent epoch of optical spectroscopy within a few months of each  {\it Chandra} observation. Details of these follow-up observations are listed in Table~3. 

	We use the Dual Imaging Spectrograph on the ARC 3.5m telescope at Apache Point Observatory to obtain a long-slit optical spectrum for each changing-look quasar. For each spectrum, we use the B400/R300 grating (spectral resolution of R $\sim$ 1000 and wavelength coverage of  $\lambda$ $\sim$ 3400 - 9200 \AA), with a $1.\!\!^{\prime\prime}5$ slit. The total exposure times range from 45 to 90 min, and the observations are taken at mean airmasses between 1.03 to 1.35, with seeing between $1.\!\!^{\prime\prime}2$ to $1.\!\!^{\prime\prime}6$. Additional spectra of spectrophotometric standard stars are obtained on each night for flux-calibration, and HeNeAr lamps are used for wavelength calibration. We reduce these optical spectra using standard \texttt{IRAF} \citep{tody86, tody93} procedures, including bias and flat-field correction, wavelength- and flux-calibration, and corrections for atmospheric extinction. Finally, we correct for Galactic extinction using the dust maps of \citet{schlegel98} and the Milky Way reddening law of \citet{cardelli89}.

	The reduced ARC 3.5m spectra are shown in Figure~\ref{fig:1}, and show that the broad emission lines and continuum emission remain faint for all of our changing-look quasars, in comparison to the SDSS spectra. This demonstrates that our changing look quasars are all still in a faint state contemporaneous with the  {\it Chandra} observations, and did not re-brighten back to a bright state. However, we opt to use the bright and faint state SDSS spectra in all our analysis (e.g. to measure $\alpha_\mathrm{OX}$) rather than these ARC 3.5m spectra, due to the superior spectrophotometric calibration of SDSS spectroscopy.

%%%%%%%%%%%%%%%%%%%%%%%%%%%%%%%%%%%%%%

\subsection{Bright state X-ray fluxes from  {\it ROSAT} and  {\it XMM-Newton}}
\label{ssc:rosat}

	We use archival X-ray observations to constrain the rest-frame $\nu L_\mathrm{2keV}$ for our sample of faded changing-look quasars in their former bright state. Although only one changing-look quasar in our sample actually has an archival  bright state X-ray detection (from {\it XMM-Newton}), we derive upper limits on $\nu L_\mathrm{2keV}$ from {\it ROSAT} for the remainder of our sample for completeness. The list of  bright state X-ray observations we use in our analysis is provided Table~4, and the derived bright state $\nu L_\mathrm{2keV}$ values are listed in Table~5.

	We first search the ESA  {\it XMM-Newton} Upper Limit Server\footnote{http://xmm.esac.esa.int/UpperLimitsServer/} and find that a deep pointed  {\it XMM-Newton} observation was obtained for J0159 from 7 July 2000, close in time to its  bright state SDSS spectrum from 23 Dec 2000. Previously, the X-ray spectrum from this {\it XMM-Newton} observation was reduced and fitted by \citet{lamassa15} to derive a 2 - 10 keV flux. We use the WebPIMMs tool to convert this 2 - 10 keV flux to a $\nu L_\mathrm{2keV}$, assuming the best-fit power-law spectral model (with $\Gamma = 2.13$ and no intrinsic absorption), and accounting for Galactic absorption. For the remaining five changing-look quasars in our sample, two (J1002 and J1021) were observed as part of the {\it XMM-Newton} Slew Survey, but were not detected. We choose not to use these upper limits from the {\it XMM-Newton} Slew Survey observations for J1002 and J1021 in our analysis, because these observations were obtained in the period between the bright and faint state SDSS spectra. Furthermore, the short exposures ($<$10 sec) in these  {\it XMM-Newton} Slew Survey observations do not provide useful upper limits on the X-ray fluxes (and are worse than upper limits from {\it ROSAT} observations). 

	For the five changing-look quasars in our sample without  {\it XMM-Newton} detections, we use archival {\it ROSAT} X-ray data to provide constraints on the  bright state X-ray flux. None of these objects were detected in any {\it ROSAT} catalogs from pointed or scan observations, and so we use processed  {\it ROSAT} All-Sky Survey images from the MPE  {\it ROSAT} Data Archive\footnote{http://www.xray.mpe.mpg.de/cgi-bin/rosat/rosat-survey} to derive upper limits. For each image, we use the SOSTA tool in the  \texttt{XIMAGE}\footnote{https://heasarc.gsfc.nasa.gov/docs/xanadu/ximage/ximage.html} software package to calculate 3$\sigma$ upper limits at the position of each changing-look quasar, which accounts for variations in the background level, effects of vignetting, and effective exposure time in the images. Our 0.1 - 2.4 keV upper limits on the {\it ROSAT}  count rates of these five changing look quasars are in the range of 0.11 - 0.42 cts s$^{-1}$, consistent with the overall estimated upper limit of 0.1 cts s$^{-1}$ for undetected sources in the {\it ROSAT} All-Sky Survey \citep{voges99}. We convert these count rates to $\nu L_\mathrm{2keV}$ using WebPIMMs, also assuming a power-law spectral model with $\Gamma= 1.8$ and accounting for Galactic absorption. These {\it ROSAT} upper limits on the bright state X-ray fluxes are also listed in Table~4. 

	We emphasize that these upper limits on the bright state X-ray fluxes from {\it ROSAT} do not provide strong constraints on the bright state $\alpha_\mathrm{OX}$. This is primarily because the {\it ROSAT} observations were obtained $\sim$10 years before the bright state SDSS spectra, and thus it is unclear if the changing-look quasars were in a bright or faint state during the {\it ROSAT} observations. Nevertheless, we derive these bright state upper limits on $\alpha_\mathrm{OX}$ for completeness, and rely instead on the XMM-COSMOS AGN sample to probe the correlation between $\alpha_\mathrm{OX}$ and $L_\mathrm{bol}/L_\mathrm{Edd}$ in more luminous AGN.

%%%%%%%%%%%%%%%%%%%%%%%%%%%%%%%%%%%%%%

\section{Modeling of SDSS optical spectroscopy}

	For the UV luminosities needed to calculate the $\alpha_\mathrm{OX}$ values of our changing-look quasars, we use their multi-epoch optical spectra from SDSS, both before and after their fading (shown in Figure~\ref{fig:1} and listed in Table~1). We first decompose the SDSS optical spectra to subtract their host galaxy starlight, using the procedure described below in Section~3.1. We then model the AGN continuum to extrapolate the power-law disk emission to obtain UV luminosities, described in Section~3.2. Finally, we model the broad emission lines in the optical spectra, described in Section~3.3.
	
\subsection{Decomposition of SDSS optical spectra \\ and host galaxy subtraction}
\label{ssc:decomposition}

	All of the analysis of our SDSS optical spectra is performed on the decomposed quasar spectra, which are the observed SDSS optical spectra after the host galaxy components have been subtracted. To perform this host galaxy subtraction, we decompose each spectrum into quasar and host galaxy components by fitting a mixture of quasar and galaxy eigenspectra \citep{vandenberk06, shen15, ruan16}. These quasar and galaxy eigenspectra are created through principal component analysis (PCA) of large samples of SDSS spectra \citep{yip04a,yip04b}. For each of our changing-look quasars, we perform this decomposition process for both its bright and faint state SDSS spectrum.

	We note that for one changing-look quasar in our sample (J2336), there are four bright state SDSS spectra available, all obtained within a period of approximately two years (MJDs of 51783, 51877, 52199, 52525). Since there is no significant variation among these four bright state spectra of J2336, we co-add them into one mean bright state spectrum to use in our spectral analysis of this object. For the MJD of this mean bright state spectrum of J2336, we simply adopt the mean MJD of 52096 from the four spectra.

	We first correct our SDSS spectra for Galactic extinction, also using the dust maps of \citet{schlegel98} and the Milky Way reddening law from \citet{cardelli89}. To enable fitting of the eigenspectra, we then resample all the SDSS spectra and eigenspectra onto a common rest-frame wavelength grid of the form log($\lambda$) = 3.35 + 0.0001$a$, for integer $a$ from 0 to 5900. This common wavelength grid is chosen to accommodate the rest-frame wavelength range of all our spectra, and is similar to the native SDSS spectral resolution. We next use the first five galaxy eigenspectra from \citet{yip04b}, and the first five quasar eigenspectra (excluding the second quasar eigenspectrum, which primarily describes the host galaxy) from \citet{yip04a}, which captures $>$98\% of the variance in SDSS quasar and galaxy spectra. We then perform a simple $\chi^2$ fit of this mixture of eigenspectra to each SDSS spectrum, with 10 total eigenspectrum amplitudes (i.e., PCA coefficients) as free parameters. In our tests, increasing the number of eigenspectra used in our decomposition to 10 quasar eigenspectra and 10 galaxy eigenspectra does not significantly improve our fits. Furthermore, we verify that using Markov Chain Monte Carlo algorithms to perform this decomposition results in the same fitting results as our simple $\chi^2$ method. However, we choose to use the computationally faster $\chi^2$ fit to enable Monte Carlo resampling of the observed spectra to robustly estimate uncertainties on all derived optical spectral properties (see Section~4.4).

	The best-fit quasar and galaxy components for both the bright and faint state SDSS spectrum from decomposition of the changing-look quasar J0126 are shown in Figure~\ref{fig:2} (top left panel), and similar figures for the remaining five objects in our sample are shown in the Appendix Figures~\ref{fig:8}-\ref{fig:12}. After subtracting the best-fit host galaxy component from the observed spectrum, we use the resultant decomposed quasar spectrum and the reported measurement uncertainties in the observed SDSS spectrum in our spectral modeling below to derive various optical properties. 

%------- FIGURE 2 -------
\begin{figure*}
\center{
\includegraphics*[width=0.47\textwidth]{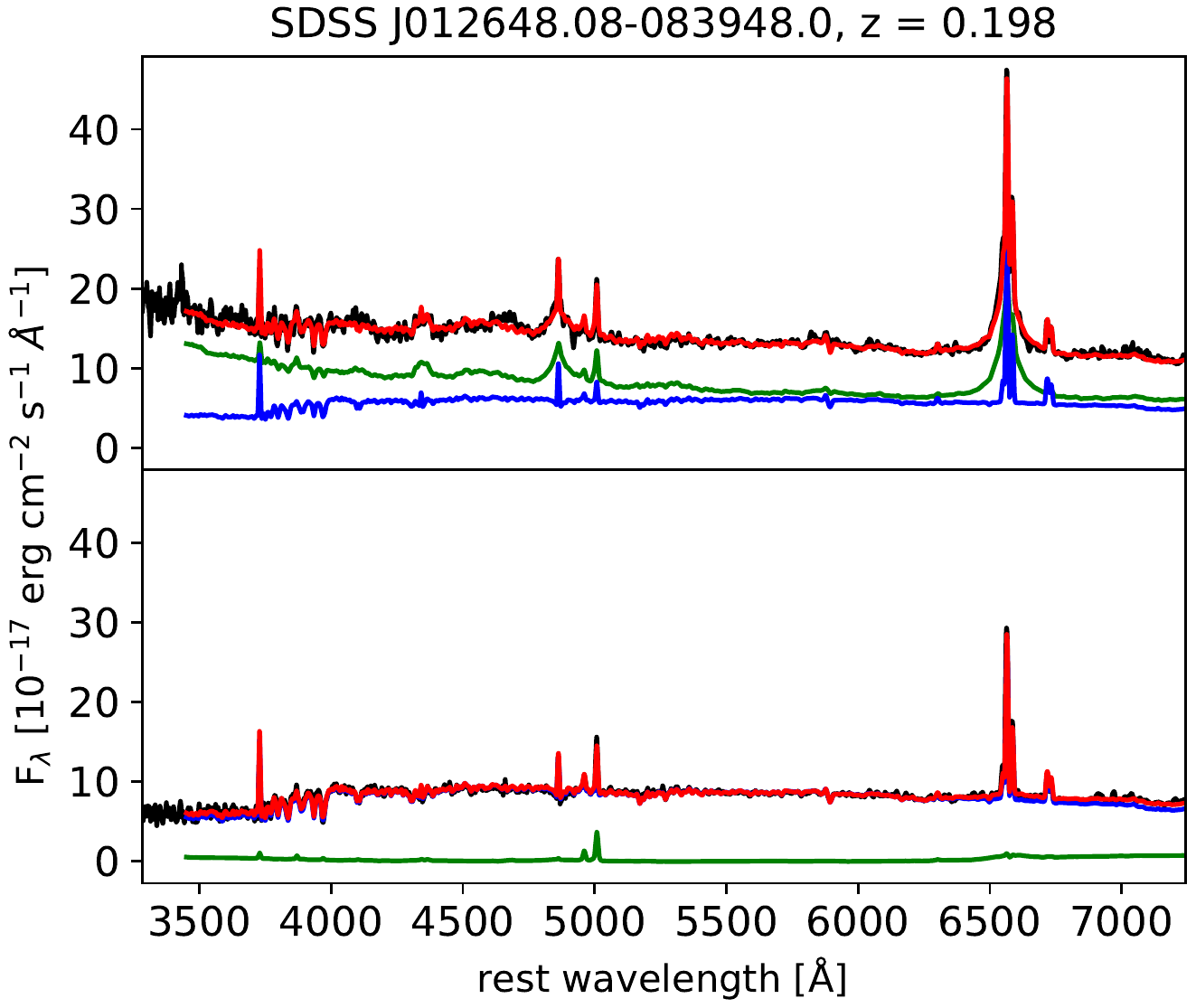} \hspace{10pt}
\includegraphics*[width=0.30\textwidth]{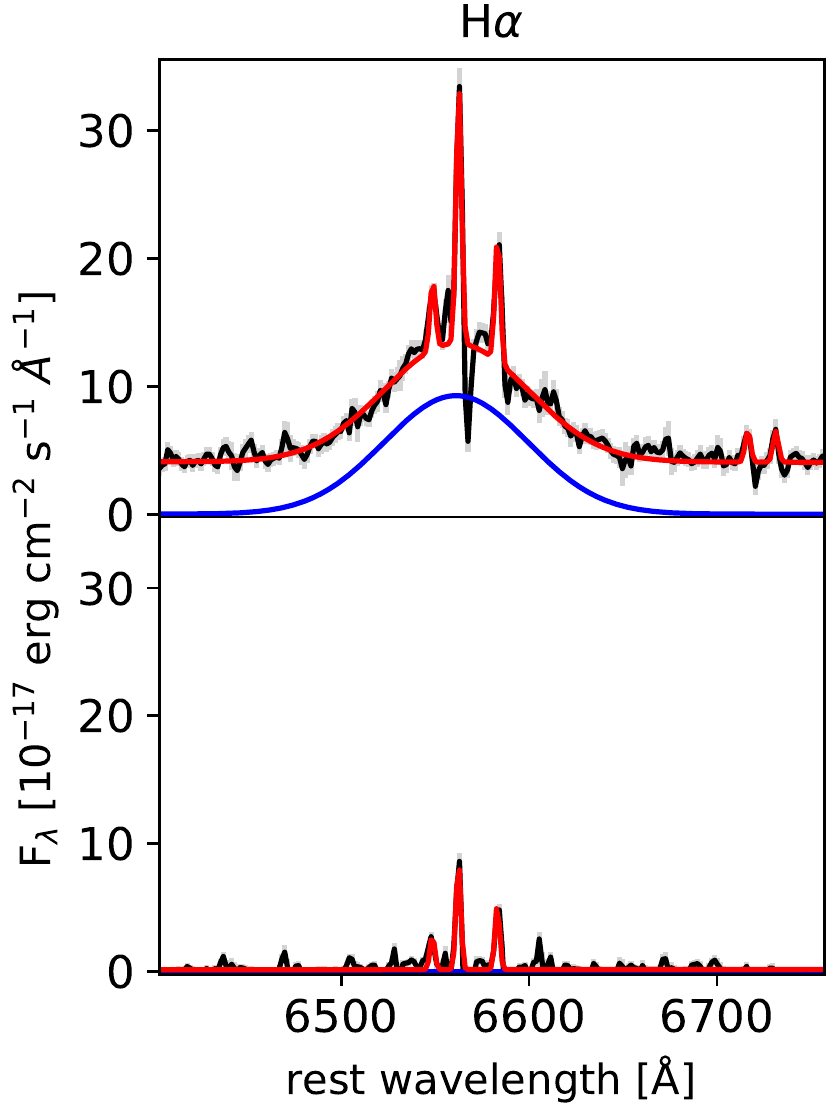}\\ \vspace{10pt}
\includegraphics*[width=0.45\textwidth]{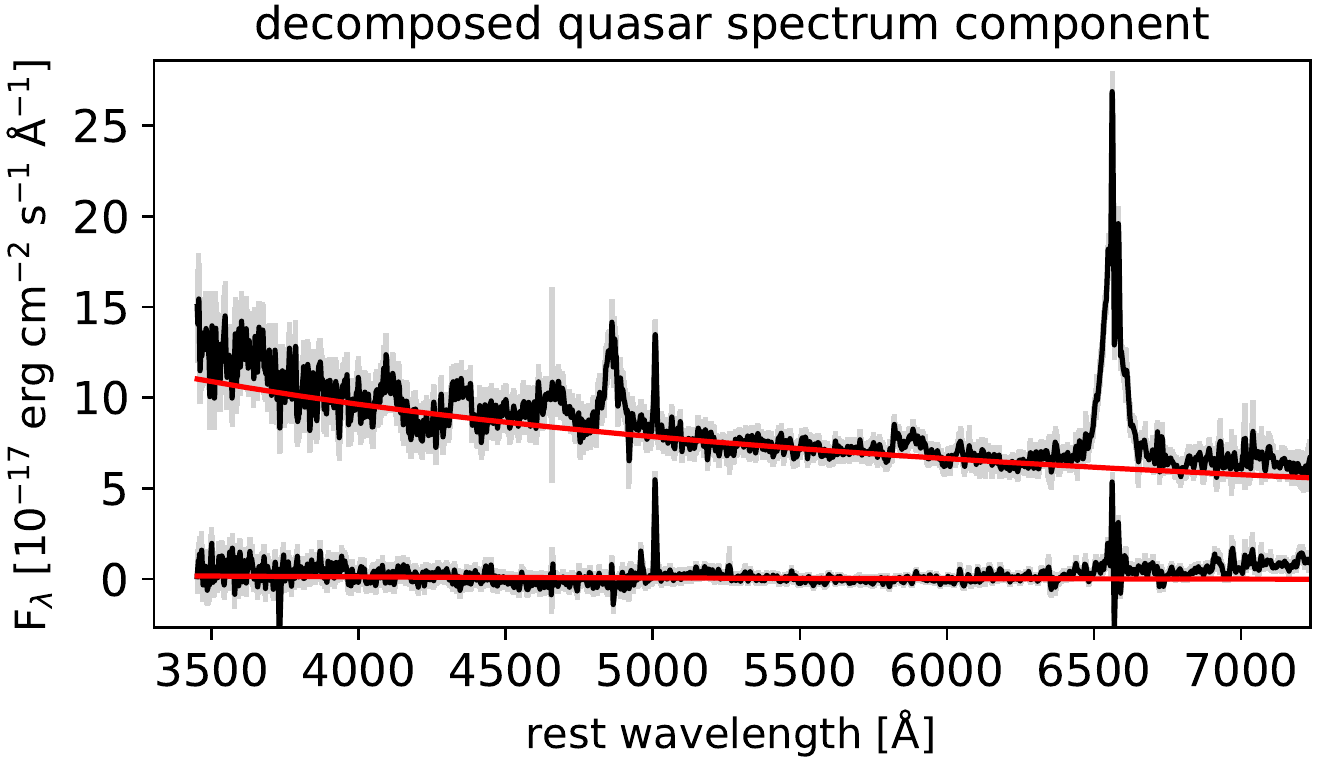}
}
\figcaption{{\bf Spectral decomposition, broad-line fitting, and continuum fitting of SDSS spectra of the changing-look quasar J0126}. In the top left figure, the observed SDSS spectra of J0126 are shown (black) during the bright state (top panel) and during its faint state (bottom panel). The best-fit quasar components (green) and host galaxy components (blue) from our spectral decomposition are also shown, along with their sum (red). The top right figure shows our spectral fitting of the H$\alpha$ region of the decomposed quasar component (black) in the bright state before its fading (top) and in the faint state after its fading (bottom). The best-fit broad H$\alpha$ emission line (blue) and full spectral model (red) are also shown. The bottom figure shows our spectral fits to the continuum emission in the decomposed quasar component (black) in both the bright and faint states. The best-fit power-law continuum (red) and the 1$\sigma$ observational uncertainties (grey) are also shown.
}
\label{fig:2}
\end{figure*}

%%%%%%%%%%%%%%%%%%%%%%%%%%%%%%%%%%%%%%
%%%%%%%%%%%%%%%%%%%%%%%%%%%%%%%%%%%%%%

\subsection{Optical AGN continuum fitting in SDSS spectra}
\label{ssc:continuum}
	
	To measure the rest-frame 2500~\AA~luminosity $\lambda L_\mathrm{2500\text{\normalfont\AA}}$ for each changing-look quasar, we use its decomposed quasar spectrum after host-galaxy subtraction to separately fit the continuum emission in both the bright and faint states. We fit the continuum emission over the rest-frame [3500, 6000] \AA~wavelength range, avoiding wavelength windows that contain prominent emission lines. Specifically, we exclude the following wavelength regions in our continuum fit: [3714, 3740]~\AA~for [O~II] $\lambda$3728; [3850, 3884]~\AA~for [Ne~III] $\lambda$3869; [4050, 4152]~\AA~for H$\delta$ $\lambda$4103; [4285, 4412]~\AA~for H$\gamma$ $\lambda$4342; [4352, 4372]~\AA~for [O~III] $\lambda$4364; [4760, 4980]~\AA~for H$\beta$ $\lambda$4863; [4945, 4972]~\AA~for [O~III]$\lambda$ 4960; [4982, 5035]~\AA~for [O~III] $\lambda$5008; and [5805, 5956]~\AA~for He~I $\lambda$5877. The wavelength ranges for these prominent emission lines are based on the composite SDSS quasar spectrum from \citet{vandenberk01}. 

	Our AGN continuum model components include a power-law, a template for optical Fe II emission from \citet{boroson92}, a model for blended high-order H Balmer broad emission lines, and a model for the Balmer continuum. At wavelengths blueward of the Balmer limit (3546 \AA), the Balmer continuum can produce a significant contribution to the observed continuum emission. We thus generate model spectra for the Balmer continuum, based on the parameterized form of \citet{grandi82} and \citet{wills85}. In this model, the Balmer continuum is an absorbed Planck function at electron temperature $T_\mathrm{e}$. This function vanishes redward of the Balmer limit, and has the form $f_\lambda = B_\lambda(T_\mathrm{e})(1 - e^{-\tau_\nu})$ blueward of the Balmer limit, where $\tau_\nu$ is the optical depth $\tau_\nu = \tau_\mathrm{0}(\lambda/\lambda_{B})^3$. We fix $T_\mathrm{e}$ to 15,000 K since variations in $T_\mathrm{e}$ within the range of 8,000 - 20,000 K have a negligible effect on the resulting spectrum in our wavelength range of interest redward of 2000~\AA. However, variations in $\tau_\mathrm{0}$ will change the slope of resultant Balmer continuum, and thus we compute Balmer continuum models over a grid of varying $\tau_\mathrm{0}$ from 0.1 - 2.0, and smooth the resulting spectra with a Gaussian filter of FWHM = 4,000 km s$^{-1}$ to account for broadening due to bulk motions of the gas.

	At wavelengths redward of the Balmer limit, high-order H Balmer emission lines blend together to produce a pseudo-continuum that should be included in our spectral fits. We thus generate model spectra for Balmer lines higher than H$\epsilon$ with upper levels up to 50, in which each line is represented by a Gaussian using their listed strengths from \citet{storey95}. In these model spectra, we assume Case B recombination, electron temperatures of 15,000 K, and electron number density of 10$^{11}$ cm$^{-3}$. For each spectrum, we smooth the width of these high-order Balmer emission lines to match the FWHM of the observed broad H$\beta$ emission line.

	We fit our continuum model to the decomposed quasar spectra in the appropriate continuum windows, using $\chi^2$ minimization. To take into account the $\tau_\mathrm{0}$ free parameter in the Balmer continuum spectral model, we perform our fit iteratively, in which we first fit the spectrum assuming an initial fixed $\tau_\mathrm{0}$, and then re-fit the spectra leaving $\tau_\mathrm{0}$ as a free parameter. The best-fitting $\tau_\mathrm{0}$ is then fixed, and this process is repeated until $\tau_\mathrm{0}$ converges. The resulting power-law continuum fits to the bright- and faint-state SDSS decomposed quasar spectra of J0126 are shown in Figure~\ref{fig:2} (bottom panel), and similar figures for the remaining five objects in our sample are shown in Appendix Figures~8-12. Based on the best-fitting power-law continuum from these fits, we obtain (1) $\lambda L_\mathrm{5100\text{\normalfont\AA}}$ for each bright state spectrum (listed in Table~1) that we will use in our  $M_\mathrm{BH}$ estimates, and (2) $\lambda L_\mathrm{2500\text{\normalfont\AA}}$ for each bright and faint state spectrum (listed in Table~5) by extrapolating their power-law continuum into the UV for use in our calculations of $\alpha_\mathrm{OX}$. In Section~\ref{ssc:verifyuv} in the Appendix, we demonstrate that this extrapolation of the optical continuum to obtain $\lambda L_\mathrm{2500\text{\normalfont\AA}}$ does not significantly affect our results. The 1$\sigma$ uncertainties on these continuum fits (including $\lambda L_\mathrm{5100\text{\normalfont\AA}}$ and $\lambda L_\mathrm{2500\text{\normalfont\AA}}$) are calculated based on Monte Carlo resampling of the SDSS spectra (see Section~4.4). 

%%%%%%%%%%%%%%%%%%%%%%%%%%%%%%%%%%%%%%
%%%%%%%%%%%%%%%%%%%%%%%%%%%%%%%%%%%%%%

\subsection{Broad emission line fitting in \\ bright state SDSS spectra}
\label{ssc:broadlines}

	To estimate the Eddington luminosity $L_\mathrm{Edd}$ for each changing-look quasar, we first fit the prominent broad H Balmer emission lines in their bright state decomposed quasar spectra to obtain viral black holes masses ($M_\mathrm{BH}$). Whenever possible, we choose to fit the broad H$\alpha$ line in the bright state decomposed quasar spectra, due to its higher signal-to-noise ratio in our spectra in comparison to other less-prominent broad emission lines. Our spectral model for H$\alpha$ includes a power-law continuum component that we fit to the continuum wavelength windows of [6400, 6500] \AA~and [6800, 7000] \AA~surrounding the H$\alpha$ line. In the H$\alpha$ line wavelength window of [6500, 6800] \AA, we include Gaussian emission line components for broad H$\alpha$, narrow H$\alpha$, the narrow [N II] $\lambda$$\lambda$6548, 6584 doublet, and the narrow [S II] $\lambda$$\lambda$6717, 6731 doublet. In our fitting, we constrain the redshifts of the narrow emission lines to be the same, and their widths to be $<$1200~km~s$^{-1}$. Furthermore, we constrain the width of the broad H$\alpha$ emission to be $>$1200~km~s$^{-1}$, while its central wavelength is left as a free parameter.

	For one quasar (J1002), H$\alpha$ is redshifted out of the wavelength range of the eigenspectra, and thus we instead fit the H$\beta$ line in the bright state decomposed quasar spectrum. The H$\beta$ spectral model for J1002 includes a power-law continuum and the optical Fe II template from \citet{boroson92} that we fit in the continuum wavelength windows of [4435, 4700] \AA~and [5100, 5535] \AA~surrounding the H$\beta$ line. In the H$\beta$ line wavelength window of [4700, 5100] \AA, we include Gaussian emission line components for broad H$\beta$, narrow H$\beta$, and the narrow [O III] $\lambda$$\lambda$4959,5007 doublet. Similar to our fitting procedure for H$\alpha$, we constrain the redshifts of the narrow lines to be the same, and their widths to be $<$1200~km~s$^{-1}$, while we constrain the width of broad H$\beta$ to be $>$1200~km~s$^{-1}$, and its central wavelength is left as a free parameter.

	We fit our broad H$\alpha$ and H$\beta$ line models to the decomposed quasar spectrum of each changing-look quasar using a simple $\chi^2$ fit. The best-fit models to both the bright- and faint-state SDSS spectra of the changing-look quasar J0126 are shown in Figure~\ref{fig:2} (top right panel), and similar figures for the remaining five objects in our sample are shown in Appendix Figures~\ref{fig:8}-\ref{fig:12}. The Full-Width at Half-Maximum (FWHM) of the broad emission components in the bright state spectra are listed in Table~1; these FWHM values are later combined with the 5100 \AA~continuum luminosities ($\lambda L_\mathrm{5100\text{\normalfont\AA}}$) to estimate $M_\mathrm{BH}$ for each changing-look quasar. The 1$\sigma$ uncertainties on these broad line fits are calculated based on Monte Carlo resampling of the SDSS spectra (see Section~4.4).
	
%%%%%%%%%%%%%%%%%%%%%%%%%%%%%%%%%%%%%%
%%%%%%%%%%%%%%%%%%%%%%%%%%%%%%%%%%%%%%

\section{Calculation of Critical Parameters}

\subsection{Spectral index $\alpha_\mathrm{OX}$ measurements}
\label{ssc:alphaOX}

	For each changing-look quasar, we calculate both a bright and faint state $\alpha_\mathrm{OX}$ \citep{tananbaum79}, which is the spectral index between the rest-frame 2500~\AA~luminosity $\lambda L_\mathrm{2500\text{\normalfont\AA}}$, and the 2~keV luminosity $\nu L_\mathrm{2keV}$. The definition of $\alpha_\mathrm{OX}$ is thus:
	
\begin{equation}
	\alpha_\mathrm{OX} = - \frac{\mathrm{log}(\lambda L_\mathrm{2500\text{\normalfont\AA}}) - \mathrm{log}(\nu L_\mathrm{2keV})}{\mathrm{log}(\nu_\mathrm{2500\text{\normalfont\AA}}) - \mathrm{log}(\nu_\mathrm{2keV})} +1.
 \end{equation}

Although the observed optical continua in our SDSS spectra probe the thin accretion disk, they do not extend blueward of rest-frame 3500~\AA, and thus $\lambda L_\mathrm{2500\text{\normalfont\AA}}$ is not directly observed. We instead estimate $\lambda L_\mathrm{2500\text{\normalfont\AA}}$ by extrapolating the best-fit power-law continuum in each spectrum from our continuum fitting (Section~3.2). The resulting bright and faint state $\lambda L_\mathrm{2500\text{\normalfont\AA}}$ and $\alpha_\mathrm{OX}$ values for each changing look quasar in our sample are listed in Table~5. The 1$\sigma$ uncertainties on our derived $\alpha_\mathrm{OX}$ values incorporate the uncertainties on $\lambda L_\mathrm{2500\text{\normalfont\AA}}$ from Monte Carlo resampling of the SDSS spectra (see Section~4.4), as well as the measurement uncertainties on $\nu L_\mathrm{2keV}$.

%%%%%%%%%%%%%%%%%%%%%%%%%%%%%%%%%%%%%%

%%%  TABLE: 5 %%%%%%%%%%%%%%%%%%%%%%%%%%%%%%%%%%%%%%

\begin{deluxetable*}{cccccc}
\tablecaption{{\bf Derived spectral properties of changing-look quasars in both their bright and faint states.} Columns include the object name, the optical luminosity state of the changing-look quasar revealed by the SDSS spectrum, bolometric Eddington ratio, UV-to-X-ray spectral index, X-ray luminosity, and UV luminosity. All uncertainties are at 1$\sigma$ confidence level, and upper/lower limits are 3$\sigma$
}
\tablehead{  
 \colhead{Object} &  \colhead{Luminosity} &  \colhead{log($L_\mathrm{bol}/L_\mathrm{Edd}$)} &  \colhead{$\alpha_\mathrm{OX}$}  &  \colhead{log($\nu L_\mathrm{2keV}$)} &  \colhead{log($\lambda L_\mathrm{2500\text{\normalfont\AA}}$)} \\ 
 \colhead{(SDSS)} &  \colhead{State} &  &  &  \colhead{[erg s$^{-1}$]}  &  \colhead{[erg s$^{-1}$]}
 }
\startdata
 \vspace{2pt}
J0126$-$0839  &  bright  &  -1.8$^{+0.1}_{-0.1}$  &  $>0.7$$^\mathrm{a}$  &  $<44.5$$^\mathrm{a}$  &  43.6$\pm$0.1 \\
 \vspace{2pt}
 &  faint  &  -3.0$^{+0.3}_{-0.3}$  &  1.1$^{+0.1}_{-0.1}$  &  41.9$^{+0.1}_{-0.1}$  & 42.3$\pm$0.1\\
 \vspace{2pt}
J0159+0033  &  bright  &  -1.6$^{+0.1}_{-0.1}$  &  1.0$^{+0.1}_{-0.1}$  &  43.8$^{+0.1}_{-0.1}$  &  43.9$\pm$0.1 \\
 \vspace{2pt}
 & faint & -2.5$^{+0.1}_{-0.1}$  &  1.0$^{+0.1}_{-0.1}$  & 42.6$^{+0.1}_{-0.1}$ & 42.4$\pm$0.1\\
 \vspace{2pt}
J1002+4509  &  bright  &  -2.0$^{+0.1}_{-0.1}$  &  $>0.7$$^\mathrm{a}$  &  $<44.9$$^\mathrm{a}$   &   44.1$\pm$0.1 \\
 \vspace{2pt}
 & faint & -2.1$^{+0.1}_{-0.1}$  &  1.0$^{+0.1}_{-0.1}$  &  43.5$^{+0.1}_{-0.1}$  & 43.5$\pm$0.1\\
 \vspace{2pt}
J1011+5442  &  bright  &  -1.8$^{+0.1}_{-0.1}$  &  $>1.0$$^\mathrm{a}$  & $<44.1$$^\mathrm{a}$  &  44.0$\pm$0.1 \\
 \vspace{2pt}
  & faint & -3.4$^{+0.4}_{-0.4}$  &  1.3$^{+0.1}_{-0.1}$  &  41.6$^{+0.3}_{-0.2}$  & 42.3$\pm$0.1 \\
 \vspace{2pt}
J1021+4645  &  bright   &  -1.8$^{+0.1}_{-0.1}$  &  $>0.8$$^\mathrm{a}$  &  $<44.4$$^\mathrm{a}$  &  43.9$\pm$0.1 \\
 \vspace{2pt}
  & faint & -2.2$^{+0.1}_{-0.1}$  &  0.9$^{+0.1}_{-0.1}$  &  43.1$^{+0.1}_{-0.1}$  & 42.8$\pm$0.1 \\
 \vspace{2pt}
J2336+0017  &  bright  &  -2.4$^{+0.2}_{-0.2}$  &  $>0.6$$^\mathrm{a}$   &  $<44.1$$^\mathrm{a}$  &  43.0$\pm$0.1 \\
 \vspace{2pt}
  & faint & -3.2$^{+0.4}_{-0.4}$  &  1.4$^{+0.1}_{-0.1}$  &  41.6$^{+0.2}_{-0.2}$  &  42.5$\pm$0.1 \\
\enddata
\tablenotetext{a}{3$\sigma$ limit} 
\label{tab:tab5}
\end{deluxetable*}

%%%%%%%%%%%%%%%%%%%%%%%%%%%%%%%%%%%%%%

\subsection{Black hole mass and \\ Eddington luminosity estimates}
\label{ssc:bhmass}

	To estimate the $M_\mathrm{BH}$ for each of our changing-look quasars, we use standard single-epoch spectroscopic $M_\mathrm{BH}$ estimation methods. In this approach, the single-epoch virial $M_\mathrm{BH}$ is based on the measured FWHM of a broad H Balmer line, as well as a radius-luminosity relation for the broad-line region from reverberation mapping of low-redshift AGNs. For the five changing-look quasars in our sample for which we measured the FWHM of the broad H$\alpha$ emission in Section~3.3, we use the relation from \citet{greene10}:
\begin{align}
M_{\rm BH, H\alpha} = (9.7 \pm 0.5)\times10^{6} \left[\frac{{\rm FWHM(H\alpha)}}{1000 {\rm \; km \; s^{-1}}}\right]^{2.06 \pm 0.06} \nonumber \\
\times \left[\frac{\lambda L_{\rm 5100\text{\normalfont\AA}}}{10^{44} {\rm \; erg \; s^{-1}}}\right]^{0.519 \pm 0.07}\; M_\odot,
\end{align}
which is based on a radius-luminosity relation from reverberation mapping by \citet{bentz09}. Similarly, for the H$\beta$ emission line in J1002, we use the relation from \citet{vestergaard06}:
\begin{align}
M_{\rm BH, H\beta} = 10^{6.91 \pm 0.02} \left[\frac{{\rm FWHM(H\beta)}}{1000 {\rm \; km \; s^{-1}}}\right]^2 \nonumber \\
 \times \left[\frac{\lambda L_{\rm 5100\text{\normalfont\AA}}}{10^{44} {\rm \; erg \; s^{-1}}}\right]^{0.5}\;M_\odot. 
\end{align}

	For each of our quasars, we calculate the Eddington luminosity using $L_\mathrm{Edd}$ = $1.26\times10^{38}$ $M_\mathrm{BH}$, for $M_\mathrm{BH}$ in units of $M_\odot$, and $L_\mathrm{bol}$ in units of erg s$^{-1}$. The 1$\sigma$ uncertainties on the $M_\mathrm{BH}$ estimates include the uncertainties on both $\lambda L_{\rm 5100\text{\normalfont\AA}}$ and broad line FWHM from Monte Carlo resampling of the SDSS spectra (see Section~4.4), as well as the uncertainties in the single-epoch virial $M_\mathrm{BH}$ relations (Equations~2 and 3), but do not account for any additional systematics. 

%%%%%%%%%%%%%%%%%%%%%%%%%%%%%%%%%%%%%%
%%%%%%%%%%%%%%%%%%%%%%%%%%%%%%%%%%%%%%

\subsection{Bolometric corrections and \\ Eddington ratio estimates}
\label{ssc:bolometriccorrections}

	To calculate the bolometric Eddington ratio $L_\mathrm{bol}/L_\mathrm{Edd}$ for each quasar, we apply an empirical bolometric correction to our observed optical and/or X-ray luminosities to estimate $L_\mathrm{bol}$. For our faint state observations, we use the bolometric correction from Equation~11 of \citet{lusso10}:
\begin{align}
	\mathrm{log}(L_\mathrm{bol}) = \mathrm{log}(\nu L_\mathrm{2-10keV}) + 1.561 \nonumber \\ - 1.853\alpha_\mathrm{OX} + 1.226\alpha_\mathrm{OX}^2.
\end{align}
This bolometric correction is dependent on $\alpha_\mathrm{OX}$, and thus takes into account the change in the shape of the SEDs for AGN at different $L_\mathrm{bol}/L_\mathrm{Edd}$. We calculate the unabsorbed $\mathrm{log}(\nu L_\mathrm{2-10keV})$ luminosity values for use in Equation~4 by using WebPIMMs to convert the observed unabsorbed 0.5 - 7 keV fluxes listed in Table~2, also assuming a power-law spectral model with $\Gamma = 1.8$.
	
	For our bright state observations, only upper limits on the bright state $\nu L_\mathrm{2keV}$ (and thus lower limits on $\alpha_\mathrm{OX}$) are available for the majority of our changing-look quasars, so we cannot use the bolometric correction in Equation~4. We instead use a bolometric correction based solely on the 3000~\AA~continuum luminosity ($\lambda L_\mathrm{3000\text{\normalfont\AA}}$) from \citet{runnoe12} for our bright state observations:
\begin{align}
	\mathrm{log}(L_\mathrm{bol}) = (0.975\pm0.028)\mathrm{log}(\lambda L_\mathrm{3000\text{\normalfont\AA}}) \nonumber \\ + (1.852\pm1.275).
\end{align}
This bolometric correction is based on the SEDs of a sample of broad-line AGN, similar to the bright states of our changing-look quasars. The $\lambda L_\mathrm{3000\text{\normalfont\AA}}$ measurements are obtained from our power-law continuum fits to the SDSS spectra (see Section~3.2). We note that for J0159 (the one changing-look quasar in our sample that has a bright state X-ray detection from {\it XMM-Newton}), the bright state log($L_\mathrm{bol}/L_\mathrm{Edd}$) calculated using Equation~4 ($-1.5\pm0.1$) and Equation~5 ($-1.6\pm0.1$) are in good agreement. In Section~\ref{ssc:verifybol} in the Appendix, we demonstrate that our results are independent of our use of these bolometric corrections.
	
%%%%%%%%%%%%%%%%%%%%%%%%%%%%%%%%%%%%%%

%------- FIGURE 3 -------
\begin{figure*} [t!]
\center{
\includegraphics[scale=0.9, angle=0.0]{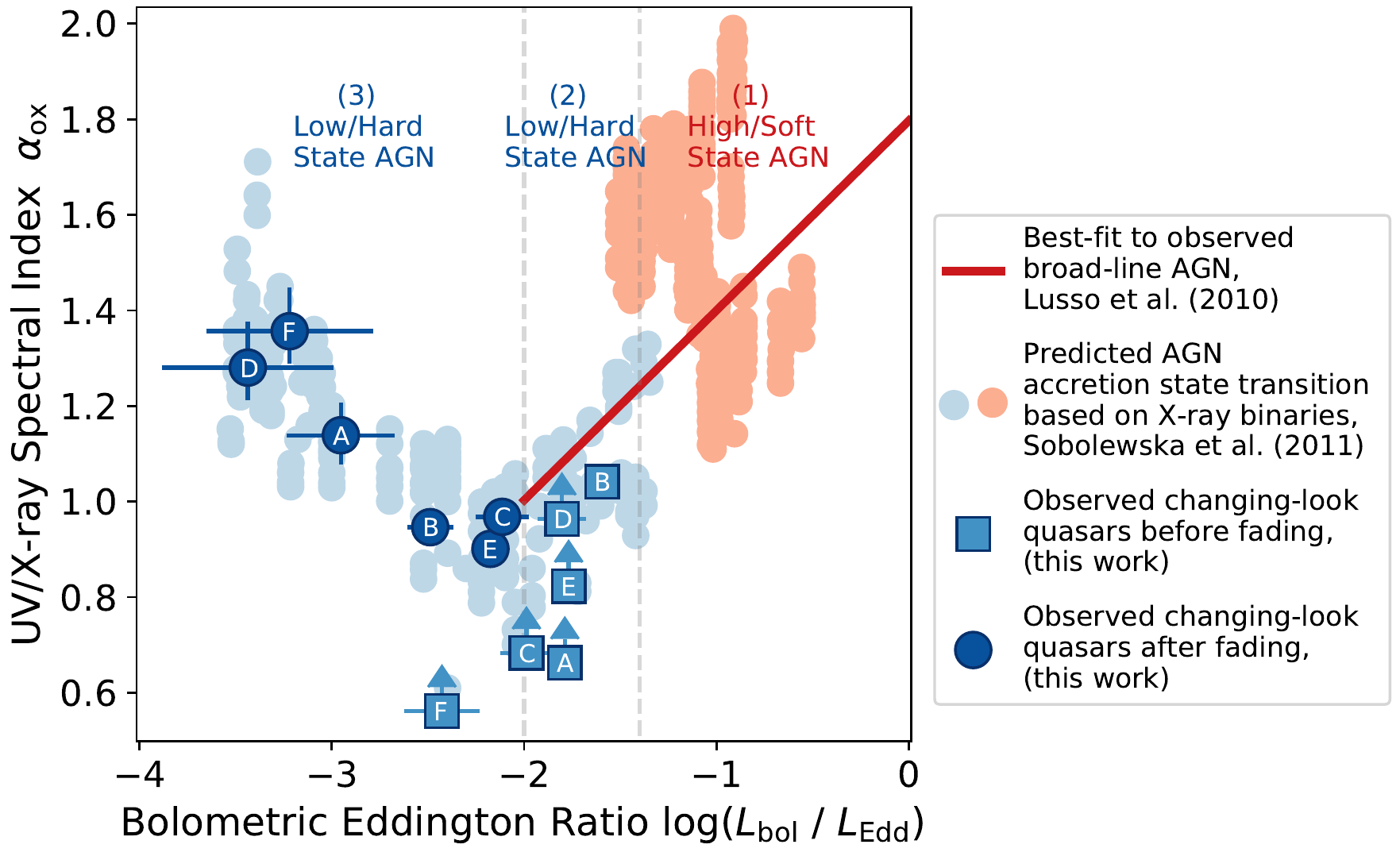}
}
\figcaption{{\bf The spectral behavior of AGN and X-ray binaries are remarkably similar across a wide range of Eddington ratios.} The UV-to-X-ray spectral index ($\alpha_\mathrm{OX}$) of changing-look quasars before their fading (light blue squares) and after their fading (dark blue circles) are shown, as well as the fit to a large sample of more luminous broad-line AGN from the XMM-COSMOS survey (red line). These observations are directly compared to predictions for accretion state transitions in AGN, based on observations of X-ray binary transitions from a high/soft state (light red circles) to a low/hard state (light blue circles). The similarity in the spectral behavior of AGN and X-ray binaries suggests that the geometries of black hole accretion flows are analogous. The lettered labels correspond to different changing-look quasars in our sample: (A) J0126, (B) J1059, (C) J1002, (D) J1011, (E) J1021, and (F) J2336.}
\label{fig:3}
\end{figure*}
% ============================

%%%%%%%%%%%%%%%%%%%%%%%%%%%%%%%%%%%%%%

\subsection{Uncertainties on derived quantities}
\label{ssc:uncertainties}
	
	We estimate the uncertainties on all quantities derived from the SDSS optical spectra (e.g., $\lambda L_\mathrm{5100\text{\normalfont\AA}}$,  $\lambda L_\mathrm{2500\text{\normalfont\AA}}$, $M_\mathrm{BH}$, $L_\mathrm{bol}/L_\mathrm{Edd}$, $\alpha_\mathrm{OX}$, etc.) through Monte Carlo resampling of the observed spectra. We first generate 1000 Monte Carlo realizations of each observed spectrum, based on the reported measurement uncertainties in each wavelength bin. For each resampled spectrum, we then repeat our spectral decomposition for host-galaxy subtraction, broad emission line fitting, and continuum fitting. The 1$\sigma$ uncertainties on the derived optical spectral quantities are thus the 1$\sigma$ spread from analysis of the 1000 resampled spectral fits.

%%%%%%%%%%%%%%%%%%%%%%%%%%%%%%%%%%%%%%
%%%%%%%%%%%%%%%%%%%%%%%%%%%%%%%%%%%%%%

\section{Results and Discussion}
\label{ssc:results}

\subsection{The observed correlation between $\alpha_\mathrm{OX}$ \\ and Eddington ratio in AGN}
\label{ssc:simulations}

	Figure~\ref{fig:3} shows the observed correlation between $\alpha_\mathrm{OX}$ and $L_\mathrm{bol}/L_\mathrm{Edd}$ for our combined AGN sample, including changing-look quasars and broad line AGN from the XMM-COSMOS survey. At Eddington ratios above $L_\mathrm{bol}/L_\mathrm{Edd} \gtrsim 10^{-2}$, the XMM-COSMOS broad line AGN display a positive correlation between $\alpha_\mathrm{OX}$ and $L_\mathrm{bol}/L_\mathrm{Edd}$ (such that $\alpha_\mathrm{OX}$ hardens as $L_\mathrm{bol}/L_\mathrm{Edd}$ decreases). This correlation is well-known and has previously been observed in many studies using single-epoch UV/X-ray observations of large samples of luminous broad-line AGN \citep[e.g.,][]{vignali03, strateva05, steffen06, just07, grupe10, jin12, wu12, vagnetti13}.  At Eddington ratios below $L_\mathrm{bol}/L_\mathrm{Edd} \lesssim 10^{-2}$, our observations of changing-look quasars in their faint state display a negative correlation between $\alpha_\mathrm{OX}$ and $L_\mathrm{bol}/L_\mathrm{Edd}$ (such that $\alpha_\mathrm{OX}$ softens as $L_\mathrm{bol}/L_\mathrm{Edd}$ decreases). Our ability to probe this correlation at $L_\mathrm{bol}/L_\mathrm{Edd} \lesssim 10^{-2}$ using faded changing-look quasars thus reveals an inversion in the correlation between $\alpha_\mathrm{OX}$ and $L_\mathrm{bol}/L_\mathrm{Edd}$ at a critical Eddington ratio of $L_\mathrm{bol}/L_\mathrm{Edd} \sim 10^{-2}$.
	
	Figure~\ref{fig:3} also compares our observed correlation between $\alpha_\mathrm{OX}$ and $L_\mathrm{bol}/L_\mathrm{Edd}$ in AGN to predictions from a X-ray binary outburst from \citet{sobolewska11a}. These predictions are based on modeling the observed X-ray spectral evolution of the X-ray binary GRO J1655-40 as it undergoes accretion state transitions during an outburst, and then scaling the evolving X-ray spectra up to AGN. This X-ray binary hosts a 6.3 $M_\odot$ black hole \citep{remillard06}, and underwent a prototypical outburst in 2005 from a low/hard state to a high/soft state, before fading back to a low/hard state. Based on X-ray spectroscopic monitoring of this outburst using the {\it Rossi X-ray Timing Explorer} (RXTE), \citet{sobolewska11a} fitted spectral models that include soft X-ray accretion disk \citep{mitsuda84} and hard X-ray Comptonized coronal components \citep{zdziarski96, coppi99}, to characterize how its X-ray spectrum changes during the outburst. To scale the observed spectral behavior of the thin disk component to AGN, \citet{sobolewska11a} assumed that the luminosity scales as $L \propto M_\mathrm{BH}$, while the temperature of the disk component follows the $T \propto M_\mathrm{BH}^{-1/4}$ scaling for Shakura-Sunyaev disks \citep{shakura73}. For the corona spectral component, it is assumed that the heating-to-cooling compactness ratio of the disk-corona system \citep{gierlinski99} changes with $L_\mathrm{bol}/L_\mathrm{Edd}$ similarly in AGN as observed in GRO J1655-40. The resulting predicted spectral behavior of accretion state transitions in AGN is then quantified by $\alpha_\mathrm{OX}$ and also shown in Figure~\ref{fig:3}, to enable direct comparisons to our observations of AGN. 

	Ideally, a direct comparison between the predicted $\alpha_\mathrm{OX}$ evolution from X-ray binaries to single-epoch observations of AGN will use an AGN sample with a narrow $M_\mathrm{BH}$ distribution. This is because if the AGN sample has a wide spread in $M_\mathrm{BH}$, the $T \propto M_\mathrm{BH}^{-1/4}$ scaling of the thin disk temperature at a fixed $L_\mathrm{bol}/L_\mathrm{Edd}$ will create an intrinsic spread in $\alpha_\mathrm{OX}$. To minimize this effect and ensure a fair comparison to predictions from X-ray binaries, the AGN sample should have a narrow range in $M_\mathrm{BH}$, and the predicted $\alpha_\mathrm{OX}$ from X-ray binaries should assume this same $M_\mathrm{BH}$ distribution. The predictions generated by \citet{sobolewska11a} in Figure~\ref{fig:3} specifically assume the narrow log-normal $M_\mathrm{BH}$ distribution from the XMM-COSMOS broad-line AGN at $L_\mathrm{bol}/L_\mathrm{Edd} \gtrsim 10^{-2}$, which have mean $\langle$log$M_\mathrm{BH}$$\rangle$ = 8.4 and standard deviation of $\sigma_{\mathrm{log}M_\mathrm{BH}}  = 0.3$ (for $M_\mathrm{BH}$ in units of $M_\odot$). Furthermore, this $M_\mathrm{BH}$ distribution is nearly identical to that of our changing-look quasars (see Table~1), which have $\langle$log$M_\mathrm{BH}$$\rangle$ = 8.3 and $\sigma_{\mathrm{log}M_\mathrm{BH}} = 0.2$. Thus, our comparison of a sample of AGN (including changing-look quasars) to predictions from X-ray binaries uses consistent and well-matched $M_\mathrm{BH}$ distributions, and can robustly probe the change in AGN SEDs as a function of  $L_\mathrm{bol}/L_\mathrm{Edd}$. This excellent match in $M_\mathrm{BH}$ distribution between the XMM-COSMOS broad-line AGN and our changing-look quasars is the reason we specifically use XMM-COSMOS AGN, instead of similar results from other previous investigations of the relation between $\alpha_\mathrm{OX}$ and  $L_\mathrm{bol}/L_\mathrm{Edd}$ in luminous quasars. We note that differences in $M_\mathrm{BH}$ between different AGN will change their $\alpha_\mathrm{OX}$ values, causing data points to spread vertically in Figure~\ref{fig:3}. Since the predictions for AGN accretion state transitions from X-ray binary outbursts by \citet{sobolewska11a} assumes a distribution in the $M_\mathrm{BH}$ of AGN, this results in the vertical `striping' pattern in their predictions as seen in Figure~\ref{fig:3}.
	
	Our observations reveal that the correlation between $\alpha_\mathrm{OX}$ and $L_\mathrm{bol}/L_\mathrm{Edd}$  in AGN is remarkably similar to that predicted from X-ray binary outbursts. Specifically, Figure~\ref{fig:3} shows that both AGN and X-ray binaries display an V-shape inversion in their spectral evolution at a critical Eddington ratio of $L_\mathrm{bol}/L_\mathrm{Edd} \sim 10^{-2}$. Since both $\alpha_\mathrm{OX}$ in AGN and $\Gamma$ in X-ray binaries probe the geometry of their disk-corona systems, their similar evolution as a function of Eddington ratio suggest that the structure of the accretion flows in AGN and X-ray binaries are analogous throughout accretions state transitions. In Section~5.2 below, we will use this result to apply theoretical models of X-ray binary accretion flows in different accretion states to explain AGN phenomenology.
	
%------- FIGURE 4 -------
\begin{figure*}[t!]
\center{
\includegraphics*[width=1.00\textwidth]{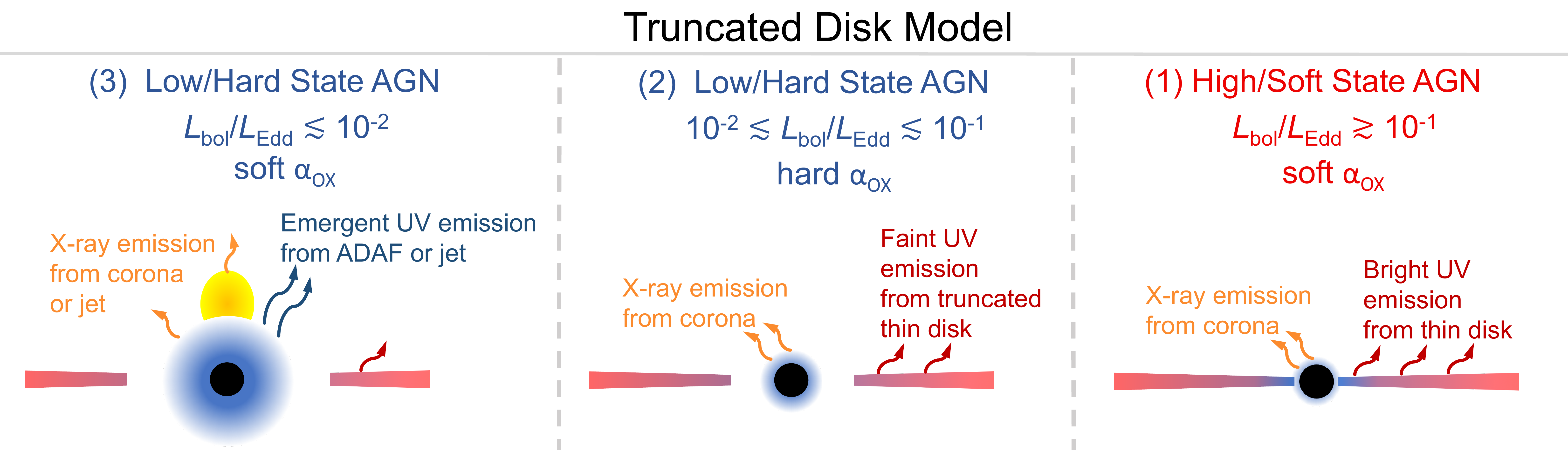}
}
\figcaption{{\bf A possible model for the geometry of AGN accretion flows in different accretion states.} The analogous geometries of black hole accretion flows enable us to use popular models of X-ray binary accretion states to interpret AGN phenomenology. The panels illustrate the application of one such X-ray binary model based on truncated disks to describe the accretion flow in fading AGN, over ranges of Eddington ratio that correspond to those labeled in Figure~\ref{fig:3}. Right panel: At high Eddington ratios ($L_\mathrm{bol}/L_\mathrm{Edd} \gtrsim 10^{-1}$) in the high/soft state, the soft $\alpha_\mathrm{OX}$ is due to bright UV emission from a thin accretion disk. Center panel: As the Eddington ratio drops ($10^{-2} \lesssim L_\mathrm{bol}/L_\mathrm{Edd} \lesssim 10^{-1}$), the inner regions of the thin disk become progressively truncated and the AGN enters the low/hard state. The truncation of the inner disk causes the UV luminosity to fade, and thus $\alpha_\mathrm{OX}$ hardens. Left panel: At low Eddington ratios  ($L_\mathrm{bol}/L_\mathrm{Edd} \lesssim 10^{-2}$), the $\alpha_\mathrm{OX}$ softens again due to the emergence of UV emission from either an advection dominated accretion flow or a jet. An alternative model without disk truncation is presented in Section~5.3.}
\label{fig:4}
\end{figure*}
%====================================
	
\subsection{Implications for the AGN/X-ray binary analogy}
	
 	Our results suggest that the geometry and overall structure of AGN accretion flows are scaled-up versions of those in X-ray binaries, which enables us to use models of X-ray binary accretion states to interpret AGN phenomenology. Figure~\ref{fig:4} illustrates the application of one such X-ray binary model to describe the evolving geometry of AGN accretion flows during state transitions, based on truncated accretion disks \citep[for a review of these and related models, see][]{done07}. AGN at Eddington ratios between $10^{-1} \lesssim L_\mathrm{bol}/L_\mathrm{Edd} \lesssim~1$ have thin accretion disks with strong thermal UV emission that result in soft $\alpha_\mathrm{OX}$; these objects represent the AGN equivalent of the high/soft state in X-ray binaries. As the Eddington ratio decreases towards $L_\mathrm{bol}/L_\mathrm{Edd} \sim 10^{-2}$, the thin disk becomes progressively truncated as the inner region evaporates into a radiatively inefficient accretion flow, leading to a decrease in UV luminosity and a hardening of $\alpha_\mathrm{OX}$; this represents the AGN transition to a low/hard state. Below $L_\mathrm{bol}/L_\mathrm{Edd} \lesssim 10^{-2}$, the hot inner portion of the disk that produces the optically-thick thermal UV emission is absent, and some other (e.g., cyclo-synchrotron or jet synchrotron) emission increasingly dominates the UV emission, causing $\alpha_\mathrm{OX}$ to soften again. These AGN are at lower Eddington ratios in the low/hard state, but above the $L_\mathrm{bol}/L_\mathrm{Edd} \lesssim10^{-5}$ regime typically associated with the quiescent accretion state in X-ray binaries.  In Section~5.3 below, we also discuss an alternative model of X-ray binary accretion state transitions, in which the thin disk is not truncated but rather displays changes in apparent temperature. 
		
	An important implication of the analogous nature of black hole accretion flows is that comparative studies can provide new insights into AGN from X-ray binaries, and vice versa. For example, observations of the spectral softening below $L_\mathrm{bol}/L_\mathrm{Edd} \lesssim 10^{-2}$ in the low/hard state of X-ray binaries have already suggested that their soft X-ray emission cannot be dominated by optically-thick thermal emission from a truncated thin disk \citep{sobolewska11b}, which can only produce spectral hardening. Instead, a new emission component must dominate the soft X-rays in this regime, such as cyclo-synchrotron emission from an advection dominated accretion flow  \citep{narayan95, wardzinski00, veledina11} or jet synchrotron emission \citep{zdziarski03, markoff04, markoff05}. Our finding of an analogous softening of $\alpha_\mathrm{OX}$ in AGN thus suggests that the faint UV/optical continuum in AGN below $L_\mathrm{bol}/L_\mathrm{Edd} \lesssim 10^{-2}$ may also be dominated by this cyclo-synchrotron or jet emission. Conversely, inconsistencies in the AGN/X-ray binary analogy can highlight differences in their accretion physics. For example, observations of X-ray binary outbursts show that decreases in Eddington ratio by factors of $\sim$100 occur on typical timescales of a few days; a simple linear scaling with black hole mass predicts similar behavior to occur in AGN on timescales of 10$^{4-5}$ years. Yet, we find here that changing-look quasars can undergo decreases in Eddington ratio by factors of $\sim$50 on timescales of $<$10 years, significantly faster than predicted. This discrepancy in variability timescales may thus support recent suggestions of the importance of magnetic pressure in shortening the timescales for changes in accretion disks around supermassive black holes \citep{noda18, dexter19}.
	
%%%%%%%%%%%%%%%%%%%%%%%%%%%%%%%%%%%%%%

%------- FIGURE 5 -------
\begin{figure*}[t!]
\center{
\includegraphics*[width=1.00\textwidth]{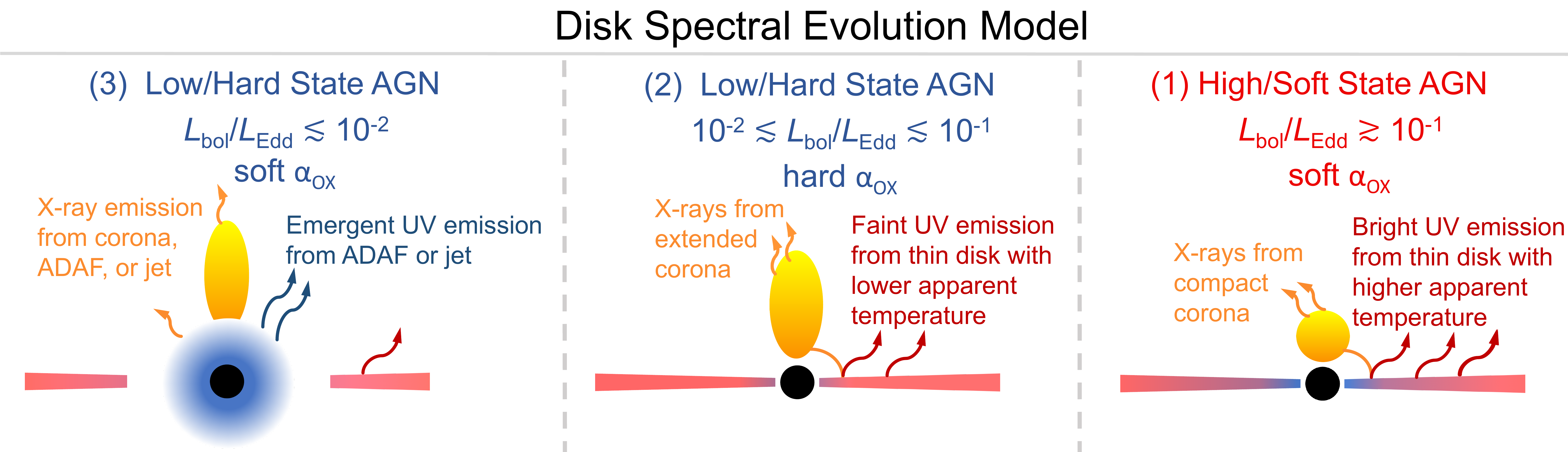}
}
\figcaption{{\bf A alternative model for the geometry of AGN accretion flows in different accretion states.}  The panels illustrate the application of a model in which the thin disk does not truncate, but rather the emergent disk spectrum decreases in apparent temperature during the transition from the high/soft state to the low/hard state. The panels display the accretion flow geometry over ranges in Eddington ratio that correspond to those labeled in Figure~\ref{fig:3}. Right panel: At high Eddington ratios ($L_\mathrm{bol}/L_\mathrm{Edd} \gtrsim 10^{-1}$) in the high/soft state, the soft $\alpha_\mathrm{OX}$ is due to bright UV emission from a thin accretion disk. Center panel: As the Eddington ratio drops ($10^{-2} \lesssim L_\mathrm{bol}/L_\mathrm{Edd} \lesssim 10^{-1}$), the inner regions of the thin disk remain near the innermost stable circular orbit, but the apparent temperature of the emergent thin disk spectrum decreases. The increasing prominence of the Componized component thus causes $\alpha_\mathrm{OX}$ to harden. Left panel: At low Eddington ratios ($L_\mathrm{bol}/L_\mathrm{Edd} \lesssim 10^{-2}$), the $\alpha_\mathrm{OX}$ softens again due to the emergence of UV emission from either an advection dominated accretion flow or a jet.}
\label{fig:5}
\end{figure*}
%====================================

%%%%%%%%%%%%%%%%%%%%%%%%%%%%%%%%%%%%%%

\subsection{Alternative models of X-ray binary state transitions}
\label{ssc:alternativemodels}

	Aside from truncated disks, our results can also be described by alternative models of X-ray binary accretion state transitions. In particular, recent observations of X-ray binaries have suggested that the thin accretion disk might not experience large-scale truncation during the transition from the high/soft state to the low/hard state \citep{reis10, dunn11, reynolds13}. Alternative models of X-ray binary accretion state transitions propose that the inner radius of the thin disk remains within a few gravitational radii of the innermost stable circular orbit, even in the low/hard state \citep{beloborodov99, markoff01, merloni02}. In a subset of these models, the observed spectral evolution during transitions from the high/soft state to the low/hard state is attributed to a decrease in the apparent temperature of the emergent thin disk spectrum, which causes a spectral hardening due to a more prominent coronal spectral component \citep{salvesen13}. This evolution of the emergent thin disk spectrum may be due to the detailed radiative transfer in the photosphere of the thin disk during a transition to the low/hard state \citep{shimura95, merloni00, davis05, davis19}, and does not involve large-scale truncation of the inner disk radius.
	
	Figure~\ref{fig:5} illustrates an application of such a disk spectral evolution model for X-ray binary accretion state transitions, to explain AGN phenomenology. In the high/soft state ($10^{-1} \lesssim L_\mathrm{bol}/L_\mathrm{Edd} \lesssim~1$), AGN emit luminous UV emission from a thin disk that extends close to the innermost stable circular orbit, which results in soft $\alpha_\mathrm{OX}$. This geometry is similar to the high/soft states in truncated disk models. However, as the Eddington ratio drops towards $L_\mathrm{bol}/L_\mathrm{Edd} \sim 10^{-2}$ and the AGN enters the low/hard state, the inner radius of the disk does not experience large-scale truncation. Instead, the luminosity fades due to a decrease in the apparent disk temperature, and the $\alpha_\mathrm{OX}$ hardens owing to a more prominent coronal component. At this point, the corona may become vertically extended as suggested by observations of an increase in the coronal variability timescales \citep{kara19}. At the lowest Eddington ratios ($L_\mathrm{bol}/L_\mathrm{Edd} \lesssim 10^{-2}$), large-scale truncation of the disk is still likely to occur as suggested by observations of X-ray binaries at low luminosities \citep{tomsick09, plant15, basak16, wangji18, vandeneijnden18}. Thus, the observed spectral softening in this regime still likely owes to synchrotron emission from a jet \citep{markoff01,markoff05} or cyclo-synchrotron emission from the advection dominated accretion flow \citep{narayan95, wardzinski00, veledina11}. 

%%%%%%%%%%%%%%%%%%%%%%%%%%%%%%%%%%%%%%%%%%%%%%%%%%%%%%

\section{Conclusion}
\label{sec:conclusion}

Although several lines of evidence have suggested links between accretion in X-ray binaries and AGN, significant uncertainties persist in whether the AGN/X-ray binary analogy holds across all accretion states and Eddington ratios. Here, we investigate how the geometry of the disk-corona system in AGN (probed by $\alpha_\mathrm{OX}$) changes as a function of Eddington ratio, and compare to observations of a prototypical X-ray binary outburst. This comparison is difficult because detailed observations of individual AGN undergoing outbursts similar to X-ray binaries are scarce, and so we instead use single-epoch observations of a {\it sample} of AGN that covers a wide range of Eddington ratios. Critically, we use faded changing-look quasars as part of our AGN sample, which enables us to robustly probe the $L_\mathrm{bol}/L_\mathrm{Edd} < 10^{-2}$ regime for the first time. Our main conclusions are the following:

\begin{itemize}
\item Fading changing-look quasars are evolving from a bright state with $L_\mathrm{bol}/L_\mathrm{Edd} \sim 10^{-1.5}$ to $10^{-2}$, into a faint state with $L_\mathrm{bol}/L_\mathrm{Edd} \sim 10^{-2}$ to $10^{-3.5}$. Although our lack of archival X-ray detections in their bright states prevents a robust measurement of their $\alpha_\mathrm{OX}$, the low Eddington ratios of $L_\mathrm{bol}/L_\mathrm{Edd} \sim 10^{-2}$ in the bright state suggest that they are likely to already be in a low/hard state even before their observed fading.

\item The observed negative correlation between $\alpha_\mathrm{OX}$ and $L_\mathrm{bol}/L_\mathrm{Edd}$ revealed by our changing-look quasars in their faint state suggests a spectral softening as AGN fade from $L_\mathrm{bol}/L_\mathrm{Edd} \sim 10^{-2}$ to lower Eddington ratios. This is in contrast to the well-known positive correlation between $\alpha_\mathrm{OX}$ and $L_\mathrm{bol}/L_\mathrm{Edd}$ in more luminous AGN at $L_\mathrm{bol}/L_\mathrm{Edd} \gtrsim 10^{-2}$. When combined, these two trends produce an inversion in the evolution of  $\alpha_\mathrm{OX}$ as a function of Eddington ratio, at a critical value of $L_\mathrm{bol}/L_\mathrm{Edd} \sim 10^{-2}$.

\item Our comparison of the observed correlation between $\alpha_\mathrm{OX}$  and $L_\mathrm{bol}/L_\mathrm{Edd}$  in AGN to predictions from a prototypical X-ray binary outburst reveals a remarkable similarity, including the inversion at the critical value of $L_\mathrm{bol}/L_\mathrm{Edd} \sim 10^{-2}$. Since these predictions are based on scaling the accretion disk and Comptonized coronal spectral components of an X-ray binary outburst to AGN, this suggests that the geometry of their disk-corona systems are analogous throughout accretion state transitions.
\end{itemize}

Looking forward, synoptic time-domain imaging surveys such as the Zwicky Transient Facility \citep{graham19b, bellm19} in the optical and eROSITA \citep{merloni12, predehl16} in the X-ray may be able to produce light curves that can follow the $\alpha_\mathrm{OX}$ evolution of {\it individual} AGN as they fade or brighten between Eddington ratios of $\sim$10$^{-4}$ to $\gtrsim$10$^{-1}$. This approach using multi-epoch UV/X-ray light curves of individual AGN (rather than inferring the properties of AGN accretion state transitions from single-epoch observations of samples of AGN) can more directly unveil the AGN analog of an X-ray binary outburst, including the inversion in $\alpha_\mathrm{OX}$ as the Eddington ratio crosses the critical value of 10$^{-2}$. Although such large changes in the Eddington ratio of AGN may occur on timescales much longer than observable, our work here already shows that factors of $\sim$50 decreases in Eddington ratio can occur on timescales of $\lesssim$10 years. More dramatic examples are thus likely to be discovered amongst the large samples of AGN monitored by these current and future surveys.

%%%%%%%%%%%%%%%%%%%%%%%%%%%%%%%%%%%%%%%%%%%%%%%%%%%%%%

\acknowledgments
J.J.R. thanks the organizers and participants of the `Unveiling the Physics Behind Extreme AGN Variability Conference' in 2017 for discussions, and Melania Nynka for guidance on reduction of the  {\it Chandra} X-ray data. 

J.J.R., S.F.A., and M.E. are supported by  {\it Chandra} Award Numbers GO7-18101A and GO8-19090A, issued by the  {\it Chandra} X-ray Observatory center, which is operated by the Smithsonian Astrophysical Observatory for and on behalf of the National Aeronautics Space Administration (NASA) under contract NAS8-03060. C.L.M, P.J.G., S.F.A., and J.J.R. are supported by the National Science Foundation under Grants No. AST-1715763 and AST-1715121. J.J.R. and D.H. acknowledge support from a Natural Sciences and Engineering Research Council of Canada (NSERC) Discovery Grant and a Fonds de recherche du Qu\'ebec-Nature et Technologies (FRQNT) Nouveaux Chercheurs Grant. J.J.R. acknowledges funding from the McGill Trottier Chair in Astrophysics and Cosmology. J.J.R. acknowledges support from the Dan David Foundation. D.H. and J.J.R. acknowledge support from the Canadian Institute for Advanced Research (CIFAR). 

The scientific results reported in this article are based to a significant degree on observations made by the  {\it Chandra} X-ray Observatory for articles by the PI team, data obtained from the  {\it Chandra} Data Archive for articles based on archival data, observations made by the  {\it Chandra} X-ray Observatory and published previously in cited articles for articles based on published results.

This work used the  {\it ROSAT} Data Archive of the Max-Planck-Institut f\"ur extraterrestrische Physik (MPE) at Garching, Germany.

This work used observations obtained with  {\it XMM-Newton}, an ESA science mission with instruments and contributions directly funded by ESA Member States and NASA.

This work used observations obtained with the Apache Point Observatory 3.5-meter telescope, which is owned and operated by the Astrophysical Research Consortium. 

Funding for the Sloan Digital Sky Survey IV has been provided by the Alfred P. Sloan Foundation, the U.S. Department of Energy Office of Science, and the Participating Institutions. SDSS-IV acknowledges support and resources from the center for High-Performance Computing at the University of Utah. The SDSS web site is www.sdss.org.

SDSS-IV is managed by the Astrophysical Research Consortium for the 
Participating Institutions of the SDSS Collaboration including the 
Brazilian Participation Group, the Carnegie Institution for Science, 
Carnegie Mellon University, the Chilean Participation Group, the French Participation Group, Harvard-Smithsonian center for Astrophysics, 
Instituto de Astrof\'isica de Canarias, The Johns Hopkins University, 
Kavli Institute for the Physics and Mathematics of the Universe (IPMU) / 
University of Tokyo, the Korean Participation Group, Lawrence Berkeley National Laboratory, 
Leibniz Institut f\"ur Astrophysik Potsdam (AIP),  
Max-Planck-Institut f\"ur Astronomie (MPIA Heidelberg), 
Max-Planck-Institut f\"ur Astrophysik (MPA Garching), 
Max-Planck-Institut f\"ur Extraterrestrische Physik (MPE), 
National Astronomical Observatories of China, New Mexico State University, 
New York University, University of Notre Dame, 
Observat\'ario Nacional / MCTI, The Ohio State University, 
Pennsylvania State University, Shanghai Astronomical Observatory, 
United Kingdom Participation Group,
Universidad Nacional Aut\'onoma de M\'exico, University of Arizona, 
University of Colorado Boulder, University of Oxford, University of Portsmouth, 
University of Utah, University of Virginia, University of Washington, University of Wisconsin, 
Vanderbilt University, and Yale University.

IRAF is distributed by the National Optical Astronomy Observatory, which is
operated by the Association of Universities for Research in Astronomy (AURA)
under a cooperative agreement with the National Science Foundation.

\facility{CXO, XMM, ROSAT, Sloan, ARC}

\software{CIAO (v.4.9; \citealt{fruscione06}), Sherpa \citep{freeman01, doe07}, WebPIMMs, XIMAGE (heasarc.gsfc.nasa.gov/docs/xanadu/ximage), IRAF \citep{tody86, tody93}}

%%%%%%%%%%%%%%%%%%%%%%%%%%%%%%%%%%%%%%%%%%%%%%%%%%%%%%

\bibliographystyle{apj}

%%%%%%%%%%%%%%%%%%%%%%%%%%%%%%%%%%

\begin{appendix}

%%%%%%%%%%%%%%%%%%%%%%%%%%%%%%%%%%%%%%

\section{Decomposition of SDSS spectra for full sample}

In Appendix Figures~\ref{fig:8}-\ref{fig:12}, we show the SDSS spectral decomposition (described in Section~3.1), broad line fitting (described in Section~3.3), and continuum power-law fitting (described in Section~3.2) results for the remaining five changing-look quasars in our sample. These figures are similar to Figure~\ref{fig:2} for J0126.

\section{Consistency checks of our results}

%%%%%%%%%%%%%%%%%%%%%%%%%%%%%%%%%%%%%%

\subsection{Verifying our extrapolated UV luminosities}
\label{ssc:verifyuv}

	In Section~3.2 above, we estimated the $\lambda L_\mathrm{2500\text{\normalfont\AA}}$ luminosities of our changing-look quasars by extrapolating the power-law AGN continuum in their SDSS spectra to 2500~\AA. Here, we demonstrate that our results are not strongly dependent on this extrapolation to obtain $\lambda L_\mathrm{2500\text{\normalfont\AA}}$. Since the observed 3500 \AA~continuum luminosity covered by our spectra also probes the thin accretion disk emission, we instead calculate an alternative $\alpha_\mathrm{OX}$$^\prime$ and $L_\mathrm{bol}/L_\mathrm{Edd}$$^\prime$, which uses the observed 3500~\AA~luminosity in Equation~1 rather than the extrapolated 2500~\AA~luminosity. Figure~\ref{fig:7} shows the behavior of this alternate $\alpha_\mathrm{OX}$$^\prime$ as a function of $L_\mathrm{bol}/L_\mathrm{Edd}$$^\prime$, similar to Figure~\ref{fig:3}. The clear softening of $\alpha_\mathrm{OX}$$^\prime$ as the Eddington ratio drops below $L_\mathrm{bol}/L_\mathrm{Edd}$$^\prime \lesssim 0.01$ is still observed, and is still a good match to predictions from X-ray binaries. This consistency of our results even when using the observed 3500~\AA~luminosity is largely because our extrapolation of the observed power-law continuum redward of 3500~\AA~to estimate $\lambda L_\mathrm{2500\text{\normalfont\AA}}$ is relatively minor, and only extends the fit over an additional 1000~\AA. Furthermore, because $\alpha_\mathrm{OX}$ is a spectral index that spans $\sim$10$^{2.8}$ Hz in frequency, small changes in the optical or X-ray luminosities do not significantly affect $\alpha_\mathrm{OX}$ (a factor of 2 change in the X-ray or UV flux will change $\alpha_\mathrm{OX}$ by only 0.11). This explicitly demonstrates that our conclusions are not strongly affected by our extrapolation of the power-law spectral continuum.

\subsection{Verifying our use of bolometric corrections}
\label{ssc:verifybol}

	In Section~4.3 above, we use a bolometric correction based on $\nu L_\mathrm{2-10keV}$ and $\alpha_\mathrm{OX}$ to estimate  $L_\mathrm{bol}$, but this bolometric correction has caveats that should be considered. We will first argue that our use of these bolometric corrections should not strongly impact our results, and then explicitly demonstrate that our conclusions are not actually dependent on the bolometric corrections. The first issue is that the faint state bolometric corrections we use are created using a sample of AGN accreting at $L_\mathrm{bol}/L_\mathrm{Edd} > 10^{-2}$, while our changing-look quasars are at lower $L_\mathrm{bol}/L_\mathrm{Edd}$ in their faint state. The faint state bolometric correction in Equation~4 is essentially a bolometric correction to $\nu L_\mathrm{2-10keV}$, but with additional terms based on $\alpha_\mathrm{OX}$ to take into account the changes in SED shape for AGN at different luminosities. For example, since the $\alpha_\mathrm{OX}$ of AGN hardens (i.e., decreases) as their X-ray luminosities decrease towards $L_\mathrm{bol}/L_\mathrm{Edd} \sim 10^{-2}$, the $\alpha_\mathrm{OX}$ terms in Equation~4 will cause the total bolometric correction to $\nu L_\mathrm{2-10keV}$ to also decrease. Since we observe an increase in $\alpha_\mathrm{OX}$ at $L_\mathrm{bol}/L_\mathrm{Edd} \lesssim 10^{-2}$ in Figure~\ref{fig:3}, applying the bolometric correction in Equation~4 to our faint state observations essentially assumes that the SEDs of AGN at $L_\mathrm{bol}/L_\mathrm{Edd} < 10^{-2}$ are similar in shape to those at $L_\mathrm{bol}/L_\mathrm{Edd} > 10^{-2}$, if their $\alpha_\mathrm{OX}$ are the same. In other words, since we observe an inversion in $\alpha_\mathrm{OX}$ at $L_\mathrm{bol}/L_\mathrm{Edd} \sim 10^{-2}$, our use of the bolometric correction in Equation~4 assumes that the SEDs of AGN at Eddington ratios of $10^{-3}$ are similar in shape to those at $10^{-1}$, since they have similar $\alpha_\mathrm{OX}$. This assumption is unlikely to significantly affect our results, since previous calculations of X-ray bolometric corrections that probe Eddington ratios down to $L_\mathrm{bol}/L_\mathrm{Edd} \lesssim 10^{-2}$ do indeed suggest that the X-ray bolometric correction has an inversion at $L_\mathrm{bol}/L_\mathrm{Edd} \sim 10^{-2}$ \cite[e.g., see Figure~5 of][]{vasudevan09}. The second issue is that the bright state bolometric correction we use in Equation~5 is based solely on $\lambda L_\mathrm{3000\text{\normalfont\AA}}$, and thus does not account for changes in SED shape as a function of $L_\mathrm{bol}/L_\mathrm{Edd}$. However, since the bulk of the multi-wavelength emission for luminous AGN is emitted in the UV, and the bright state bolometric correction we used is based on $\lambda L_\mathrm{3000\text{\normalfont\AA}}$, changes in the X-ray emission as a function of $L_\mathrm{bol}/L_\mathrm{Edd}$ will not strongly affect this bolometric correction in this regime. 

\begin{figure*}
\center{
\includegraphics*[width=0.65\textwidth]{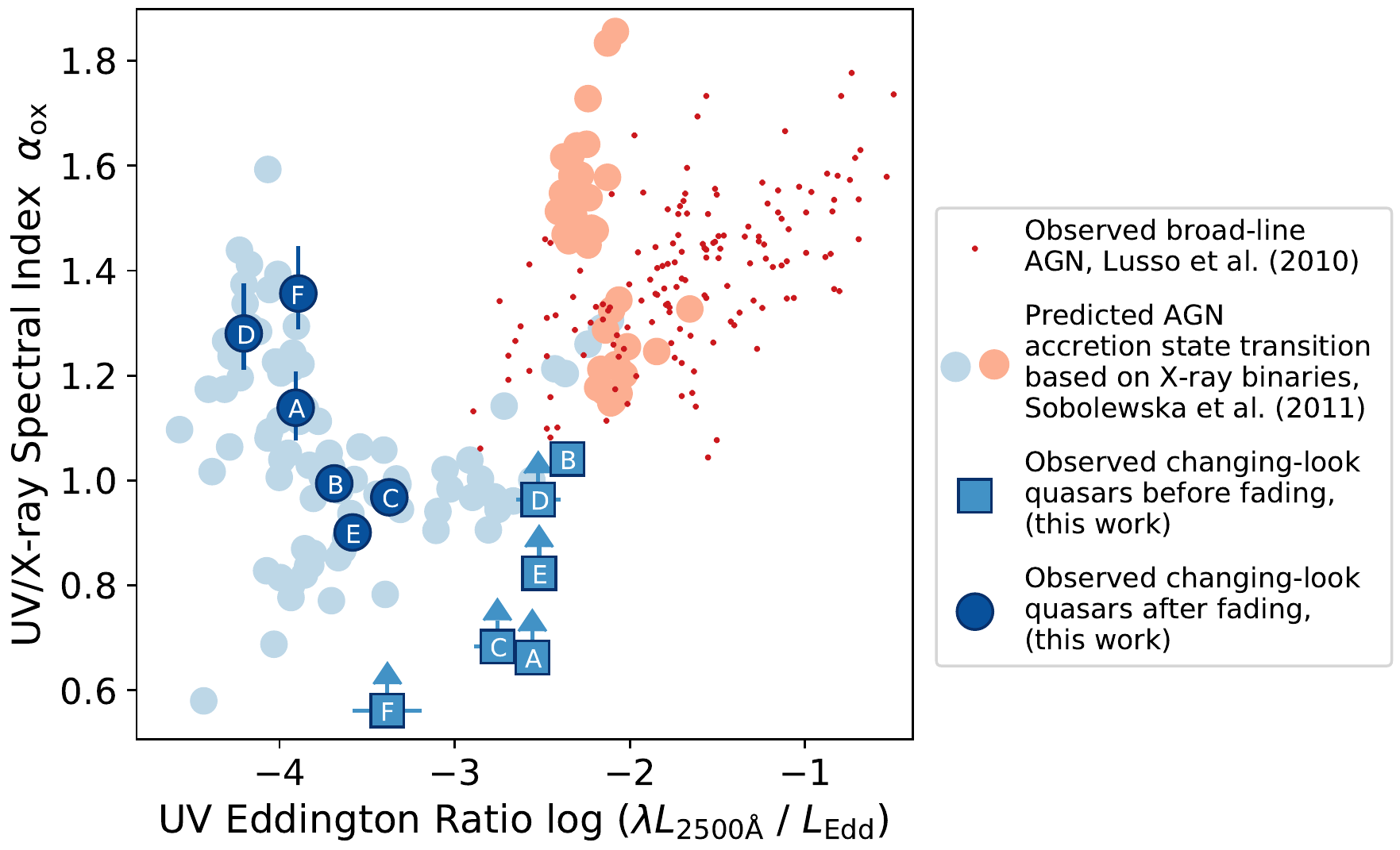}
}
\figcaption{{\bf The observed $\alpha_\mathrm{OX}$ behavior of AGN as a function of UV Eddington ratio.} This figure is similar to Figure~\ref{fig:3}, except that we show the UV Eddington ratio, which is based on the UV luminosity rather than the bolometric luminosity (i.e., we do not make a bolometric correction). The predicted inversion in the correlation between $\alpha_\mathrm{OX}$ and UV Eddington ratio is still clearly observed, and the changing-look quasar observations are still an excellent match to predictions from X-ray binary transitions. This demonstrates that our results are independent of the bolometric corrections we use.
}
\label{fig:6}
\end{figure*}

\begin{figure*}
\center{
\includegraphics*[width=0.65\textwidth]{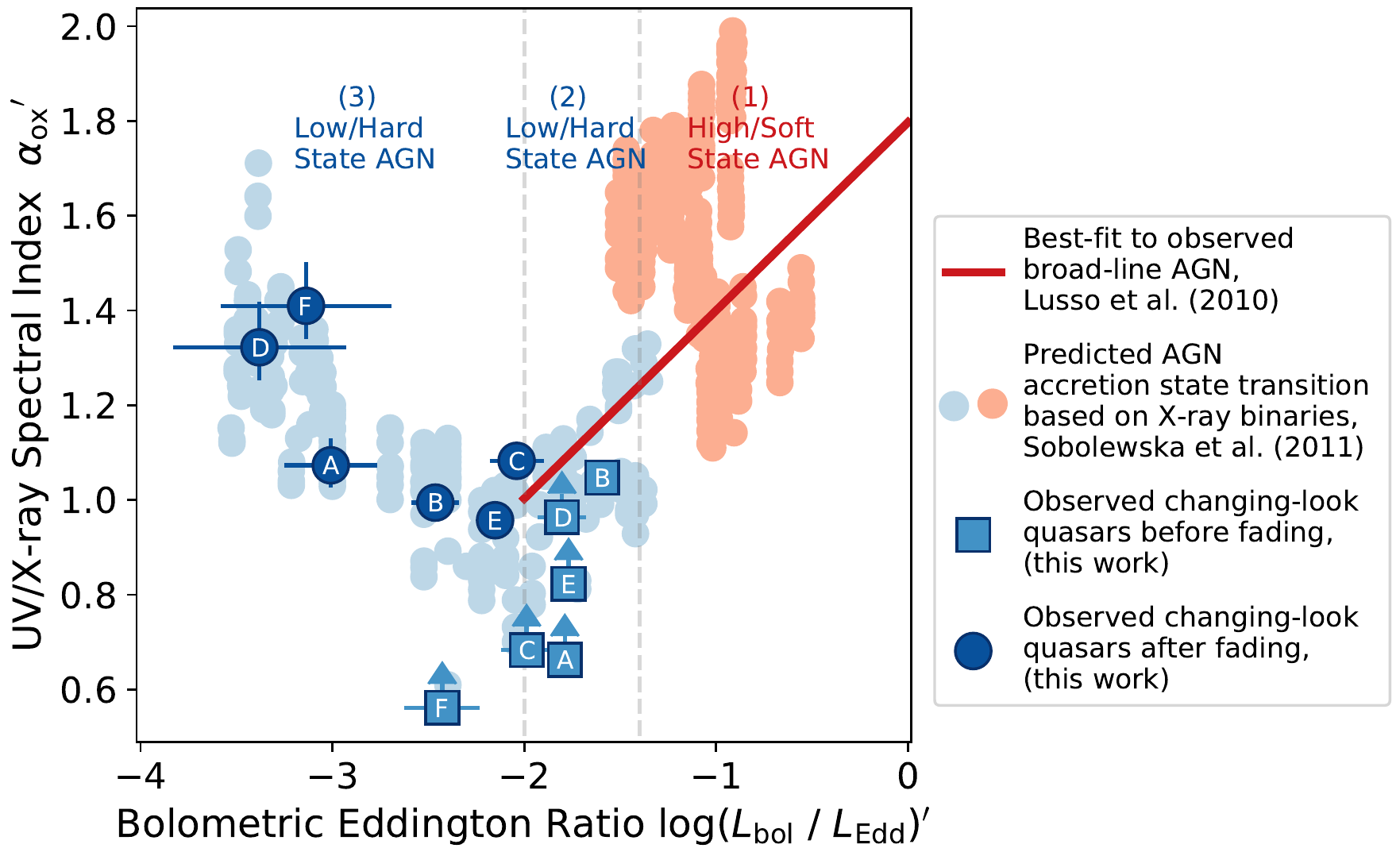}
}
\figcaption{{\bf The observed $\alpha_\mathrm{OX}$$^\prime$ behavior of AGN as a function of Eddington ratio $L_\mathrm{bol}/L_\mathrm{Edd}$$^\prime$, both calculated using $\lambda L_\mathrm{3500\text{\normalfont\AA}}$}. This figure is exactly the same as Figure~\ref{fig:2}, except that we have used the observed $\lambda L_\mathrm{3500\text{\normalfont\AA}}$ to calculate $\alpha_\mathrm{OX}$ and $L_\mathrm{bol}/L_\mathrm{Edd}$ for the changing-look quasars, instead of the extrapolated $\lambda L_\mathrm{2500\text{\normalfont\AA}}$. The softening of $\alpha_\mathrm{OX}$ at $L_\mathrm{bol}/L_\mathrm{Edd} \lesssim 10^{-2}$ is still clearly observed, and the changing-look quasar observations are still a good match to predictions from X-ray binary outbursts. This demonstrates that our results are not strongly dependent on our extrapolation of the power-law continuum in the optical spectra to 2500~\AA.
}
\label{fig:7}
\end{figure*}

	We next explicitly demonstrate that our conclusions are not actually dependent on the bolometric corrections. In Figure~\ref{fig:6}, we show the $\alpha_\mathrm{OX}$ behavior of our changing-look quasars as a function of the UV Eddington ratio $\lambda L_\mathrm{2500\text{\normalfont\AA}}/L_\mathrm{Edd}$ (i.e., without making any bolometric corrections), in comparison to predictions for accretion state transitions for a 10$^8$~$M_\odot$ AGN from Figure~1 (right panel) of \citet{sobolewska11a}. The inversion in the evolution of $\alpha_\mathrm{OX}$ is still clearly observed, and the changes in $\alpha_\mathrm{OX}$ remain in good agreement with predictions based on X-ray binaries. 

%%%%%%%%%%%%%%%%%%%%%%%%%%%%%%%%%%%%%%

%------- FIGURE 6 -------

%------- FIGURE 7 -------

%%%  APPENDIX FIG: 8 %%%%%%%%%%%%%%%%%%%%%%%%%%%%%%%%%%%%%%
\begin{figure*}
\center{
\includegraphics*[width=0.47\textwidth]{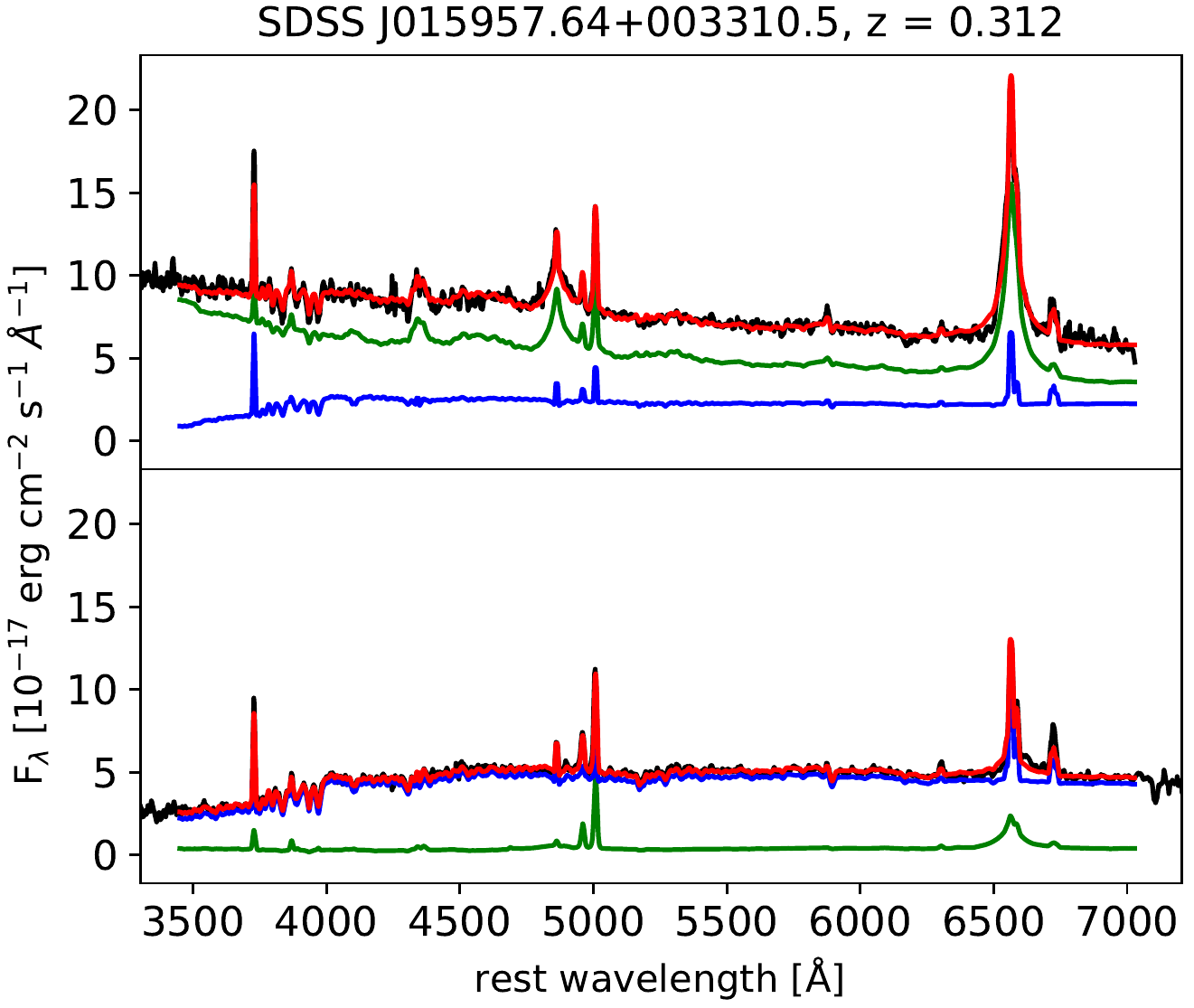} \hspace{10pt}
\includegraphics*[width=0.30\textwidth]{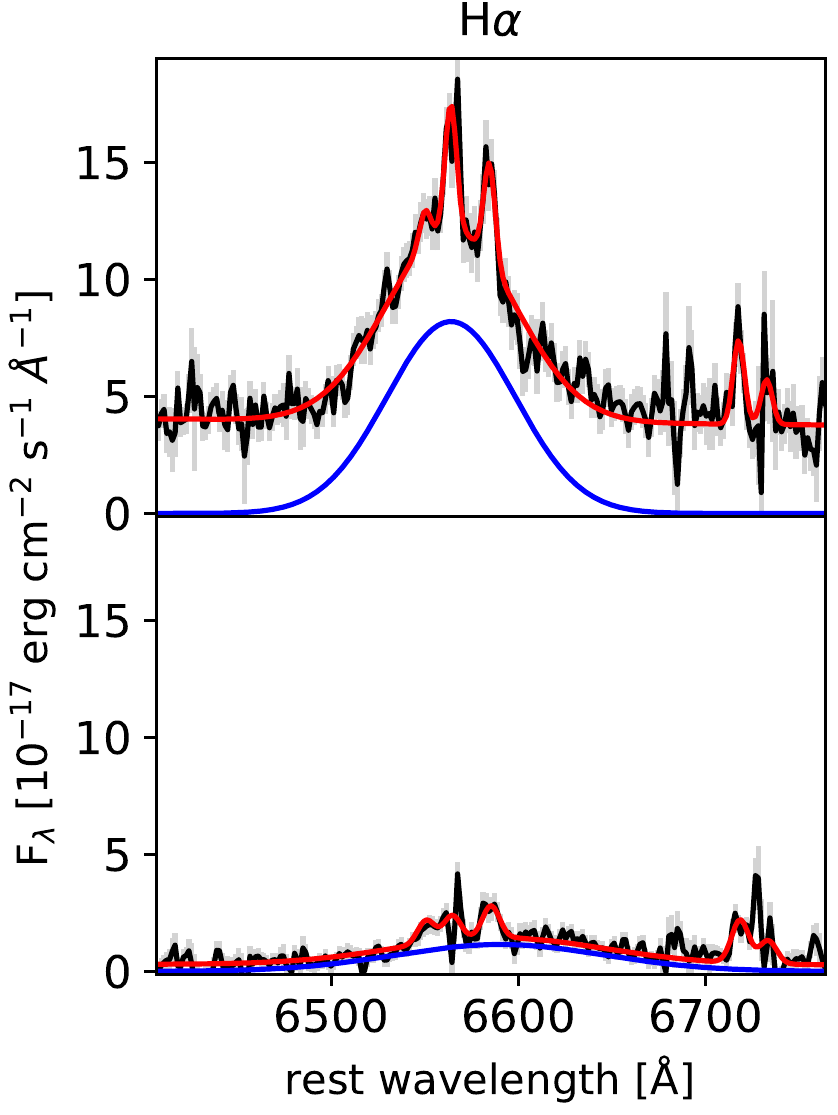}\\ \vspace{10pt}
\includegraphics*[width=0.45\textwidth]{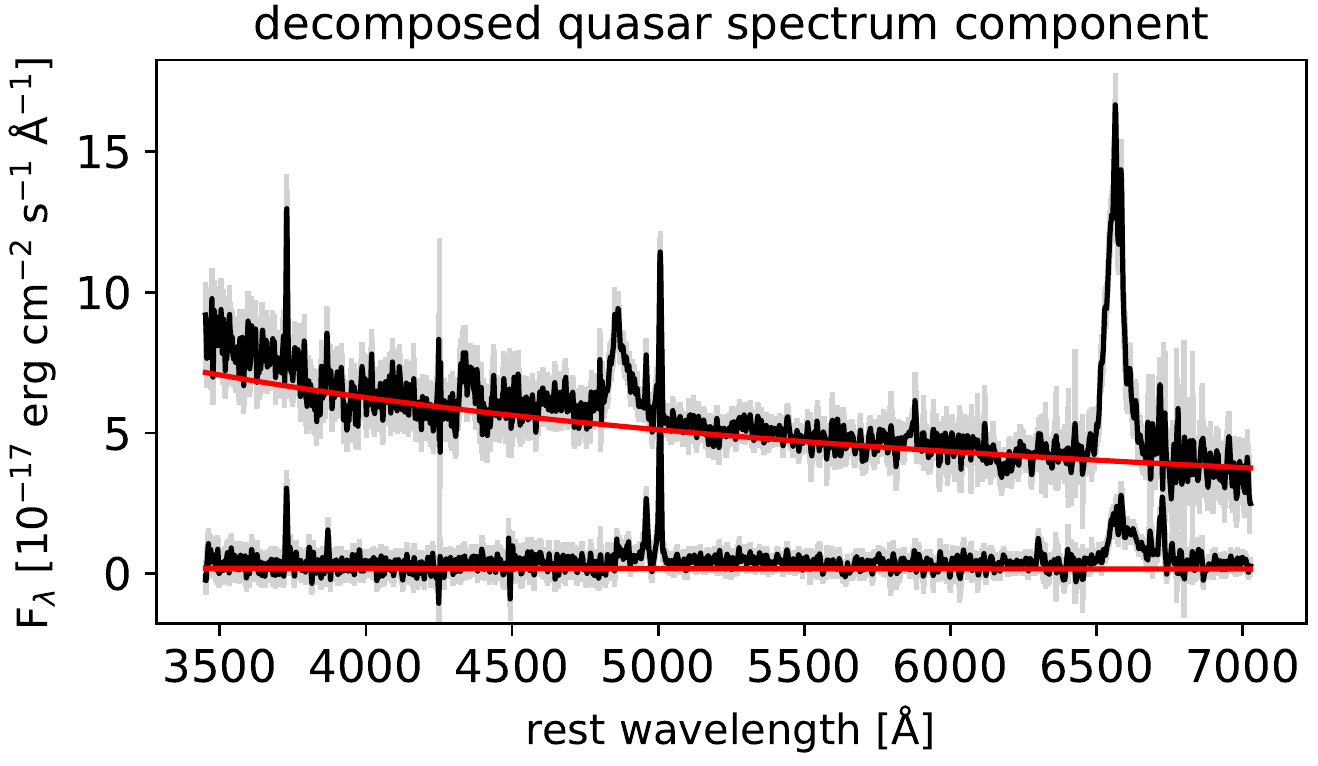}
}
\figcaption{{\bf Spectral decomposition, broad-line fitting, and continuum fitting of SDSS spectra of the changing-look quasar J0159}. 
}
\label{fig:8}
\end{figure*}

%%%  APPENDIX FIG: 9 %%%%%%%%%%%%%%%%%%%%%%%%%%%%%%%%%%%%%%
\begin{figure*}
\center{
\includegraphics*[width=0.47\textwidth]{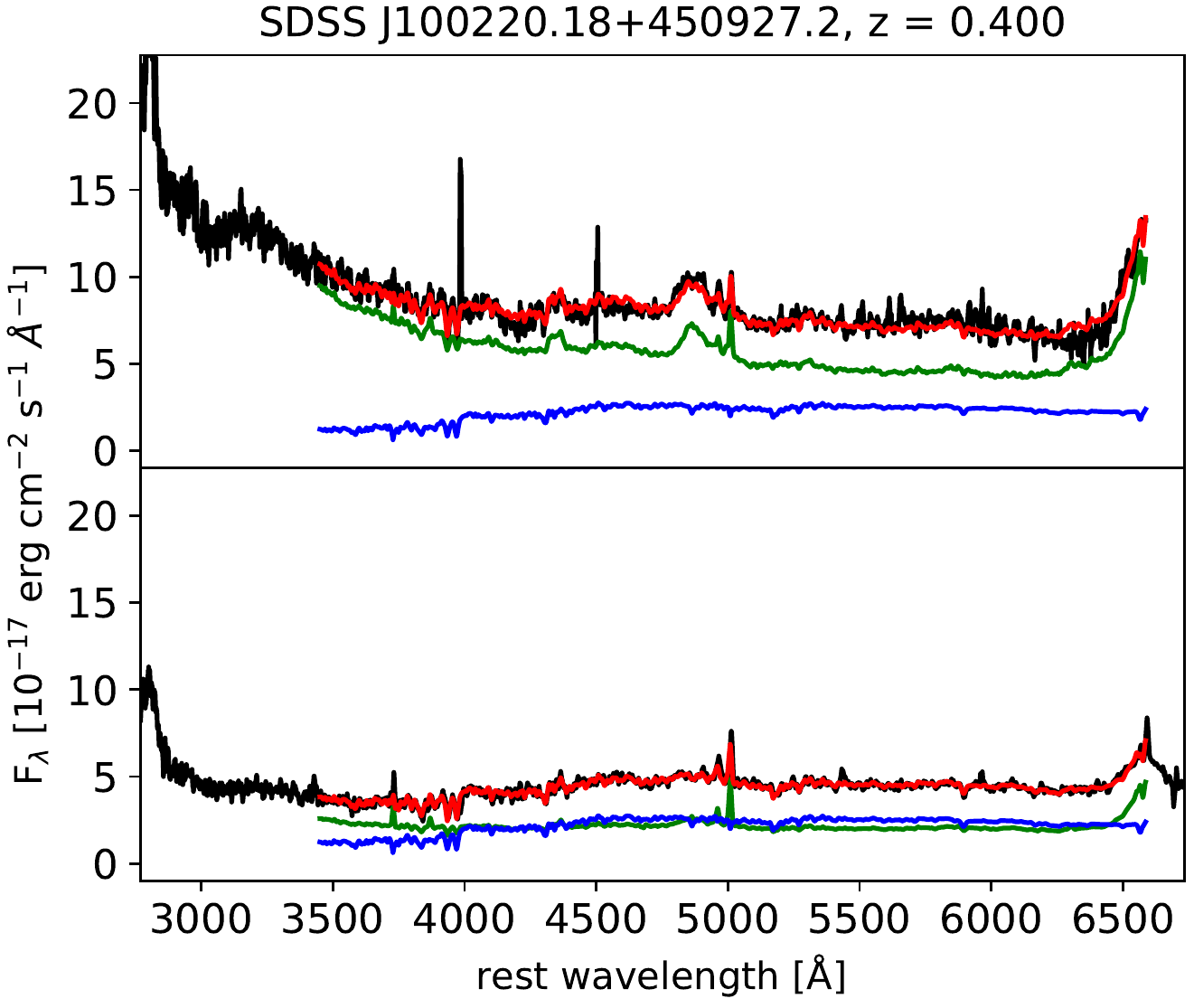} \hspace{10pt}
\includegraphics*[width=0.30\textwidth]{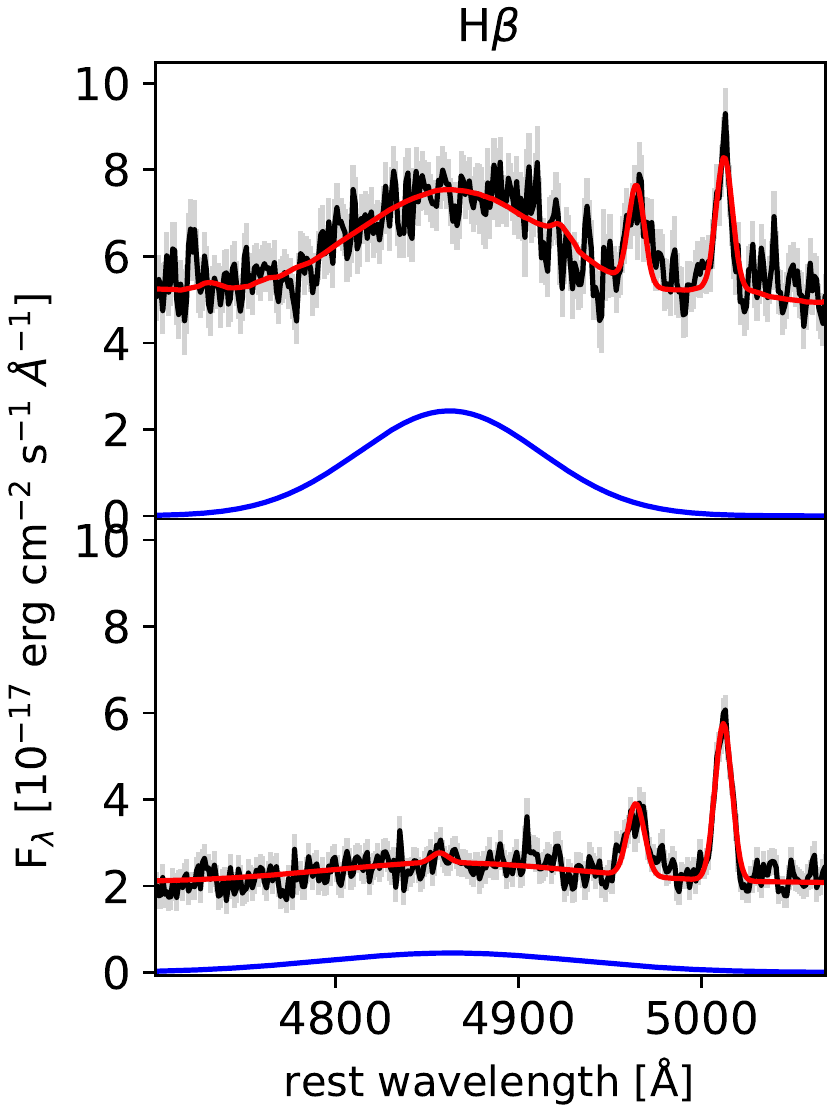}\\ \vspace{10pt}
\includegraphics*[width=0.45\textwidth]{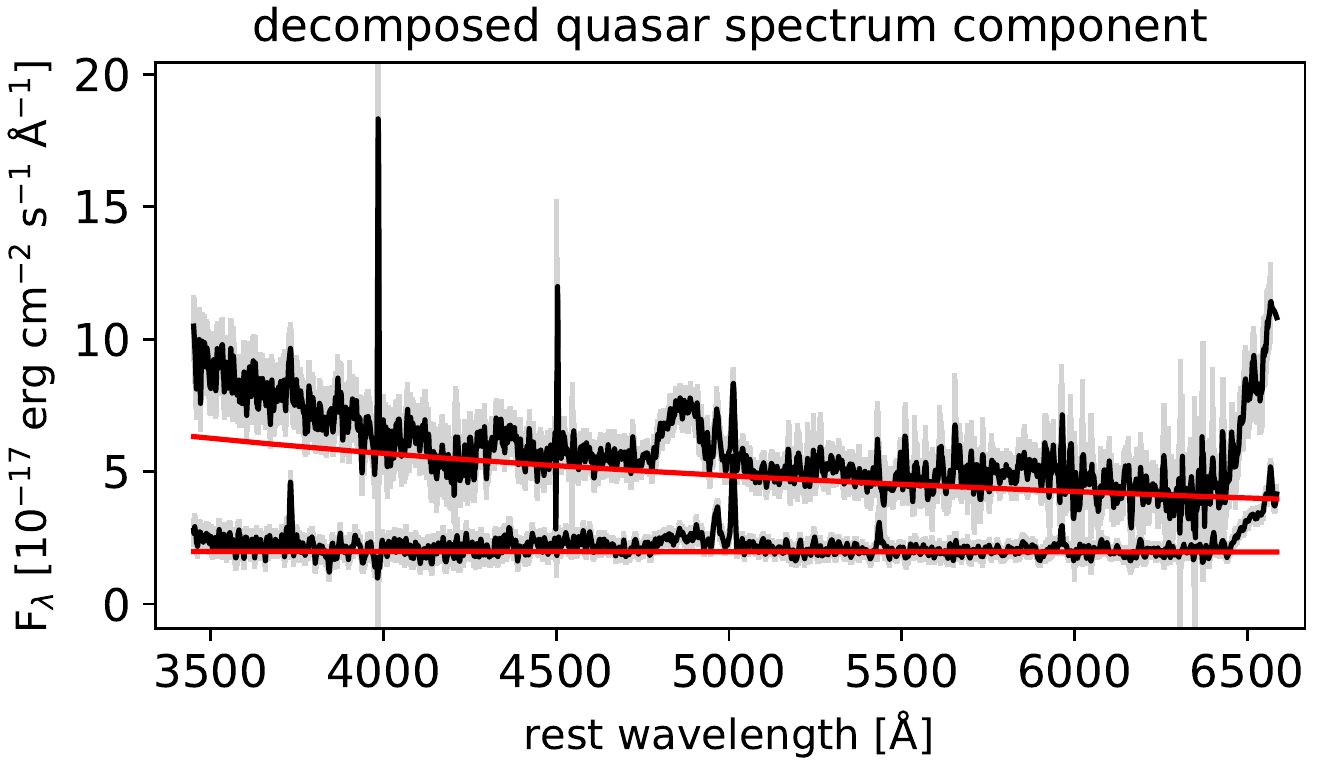}
}
\figcaption{{\bf Spectral decomposition, broad-line fitting, and continuum fitting of SDSS spectra of the changing-look quasar J1002}. 
}
\label{fig:9}
\end{figure*}

%%%  APPENDIX FIG: 10 %%%%%%%%%%%%%%%%%%%%%%%%%%%%%%%%%%%%%%
\begin{figure*}
\center{
\includegraphics*[width=0.47\textwidth]{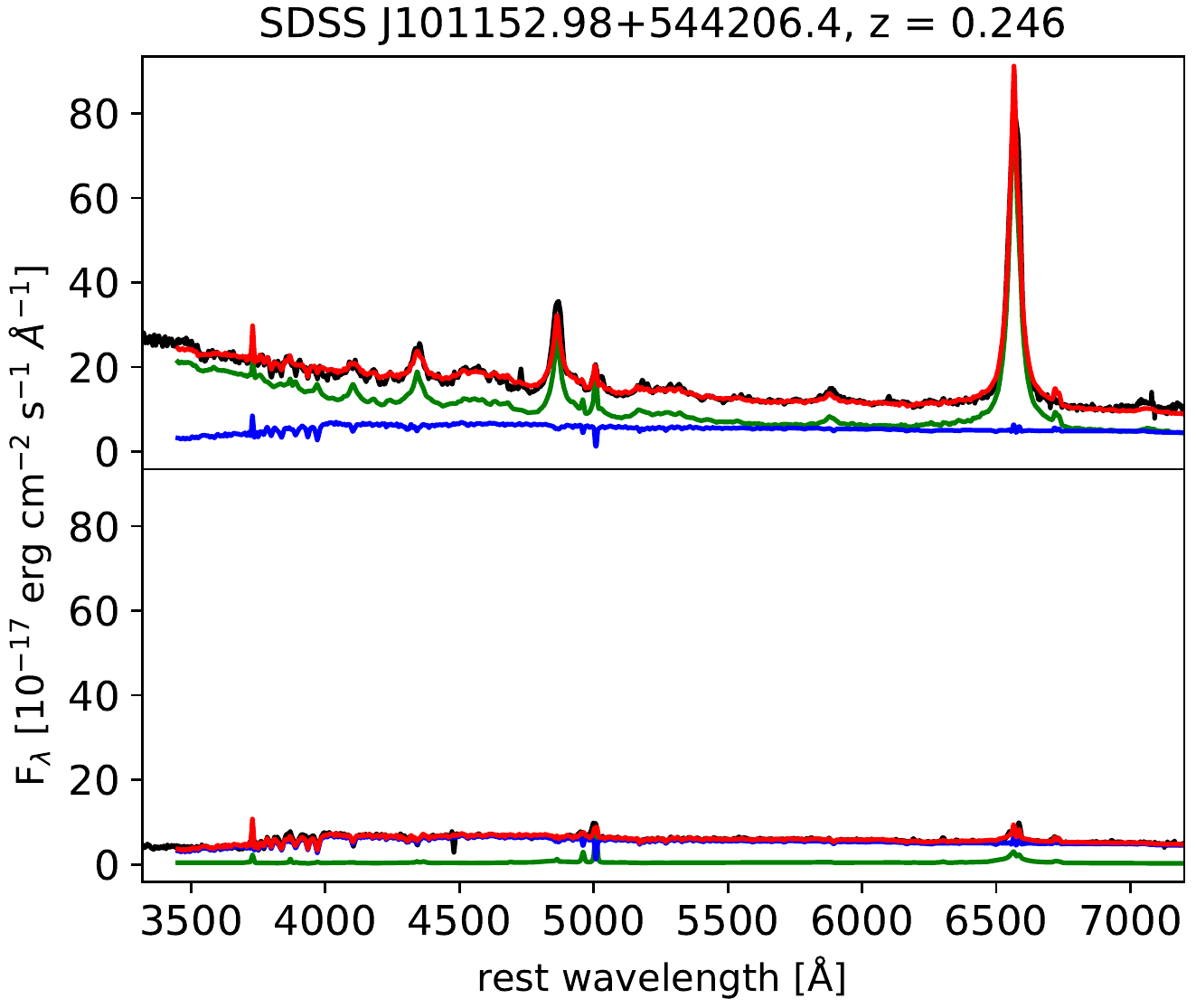} \hspace{10pt}
\includegraphics*[width=0.30\textwidth]{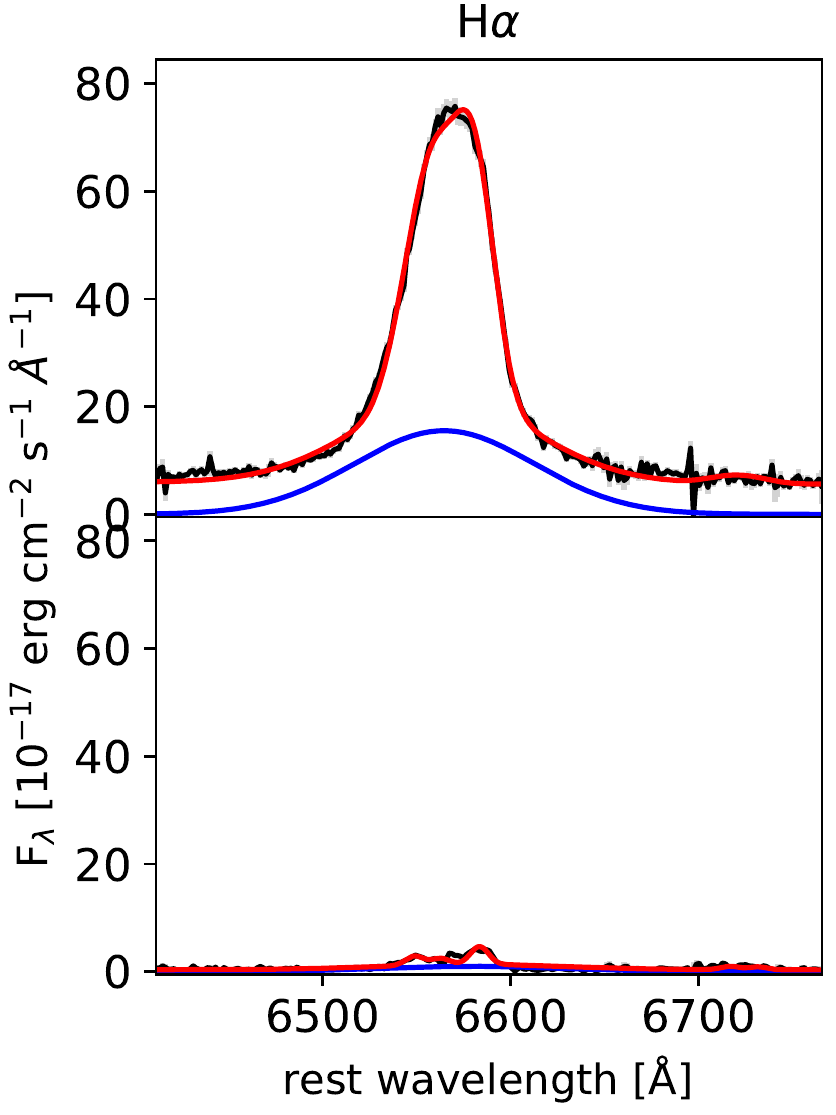}\\ \vspace{10pt}
\includegraphics*[width=0.45\textwidth]{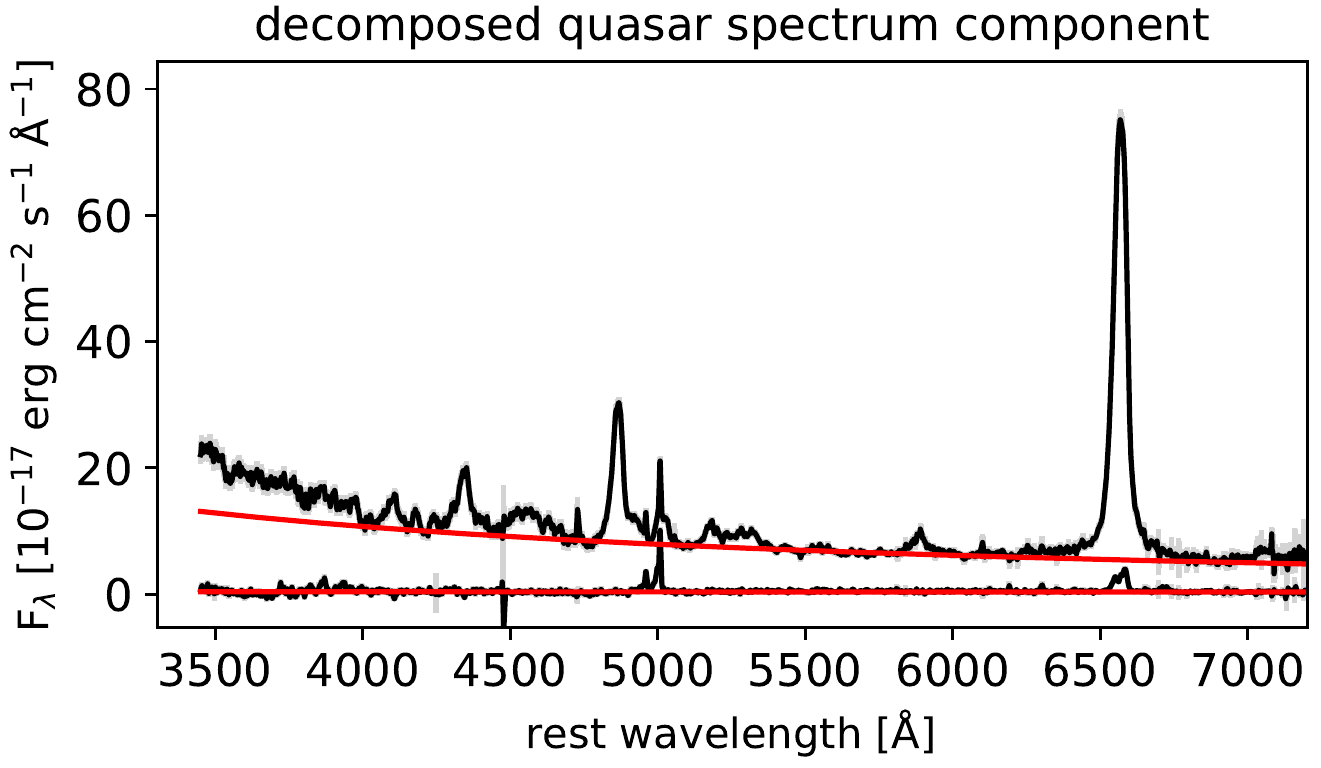}
}
\figcaption{{\bf Spectral decomposition, broad-line fitting, and continuum fitting of SDSS spectra of the changing-look quasar J1011}. 
}
\label{fig:10}
\end{figure*}

%%%  APPENDIX FIG: 11 %%%%%%%%%%%%%%%%%%%%%%%%%%%%%%%%%%%%%%
\begin{figure*}
\center{
\includegraphics*[width=0.47\textwidth]{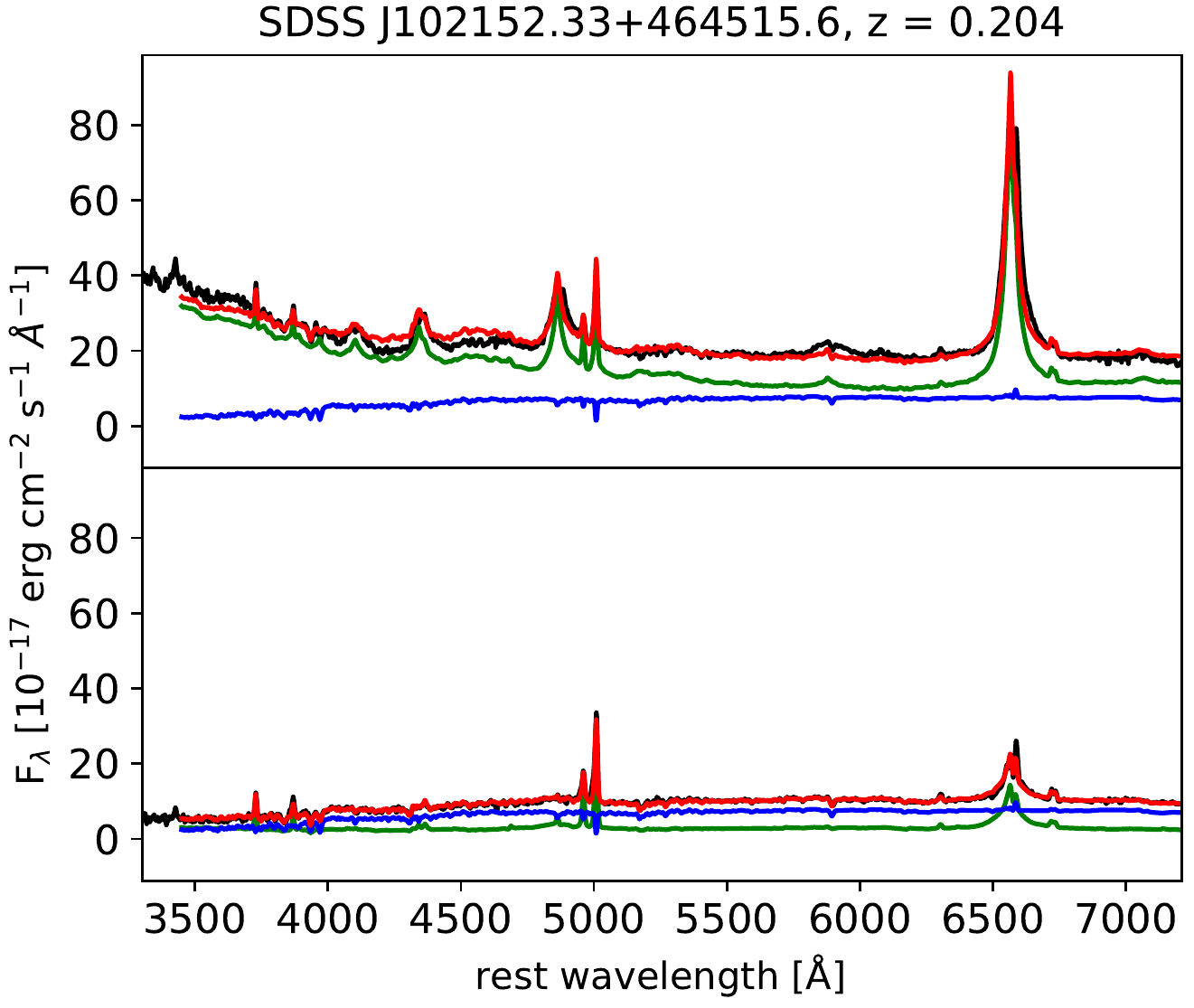} \hspace{10pt}
\includegraphics*[width=0.30\textwidth]{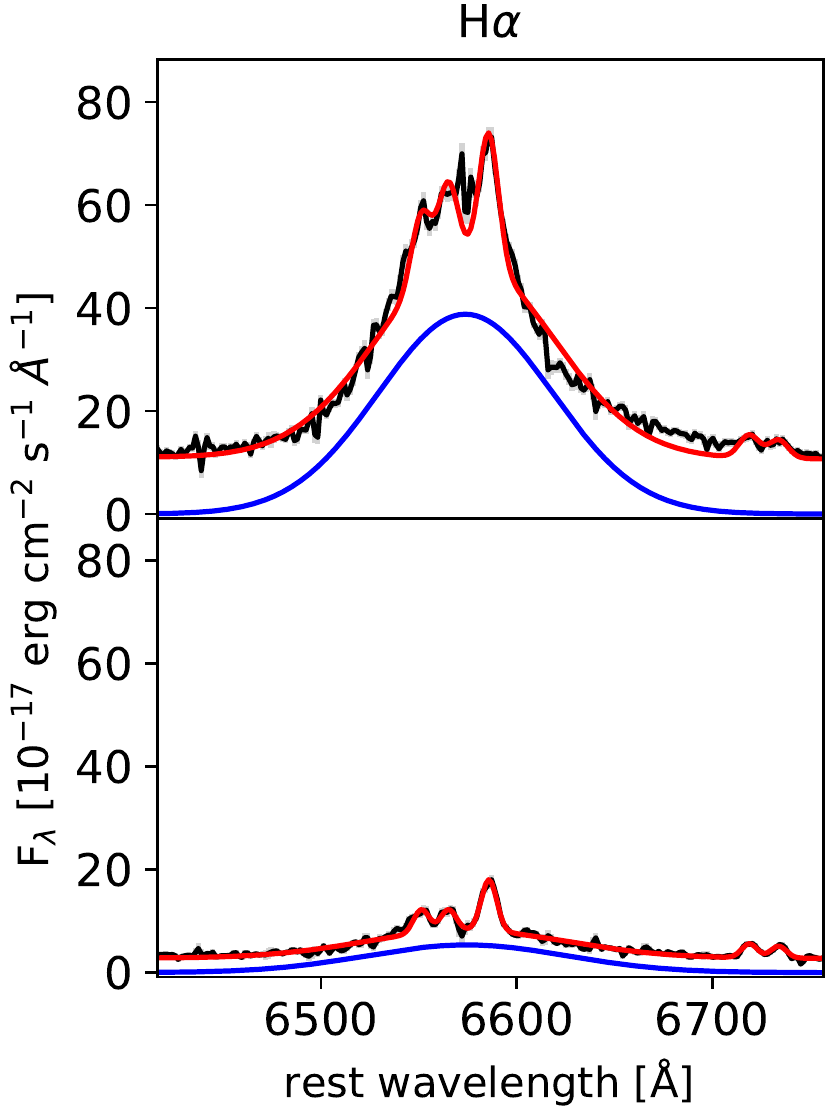}\\ \vspace{10pt}
\includegraphics*[width=0.45\textwidth]{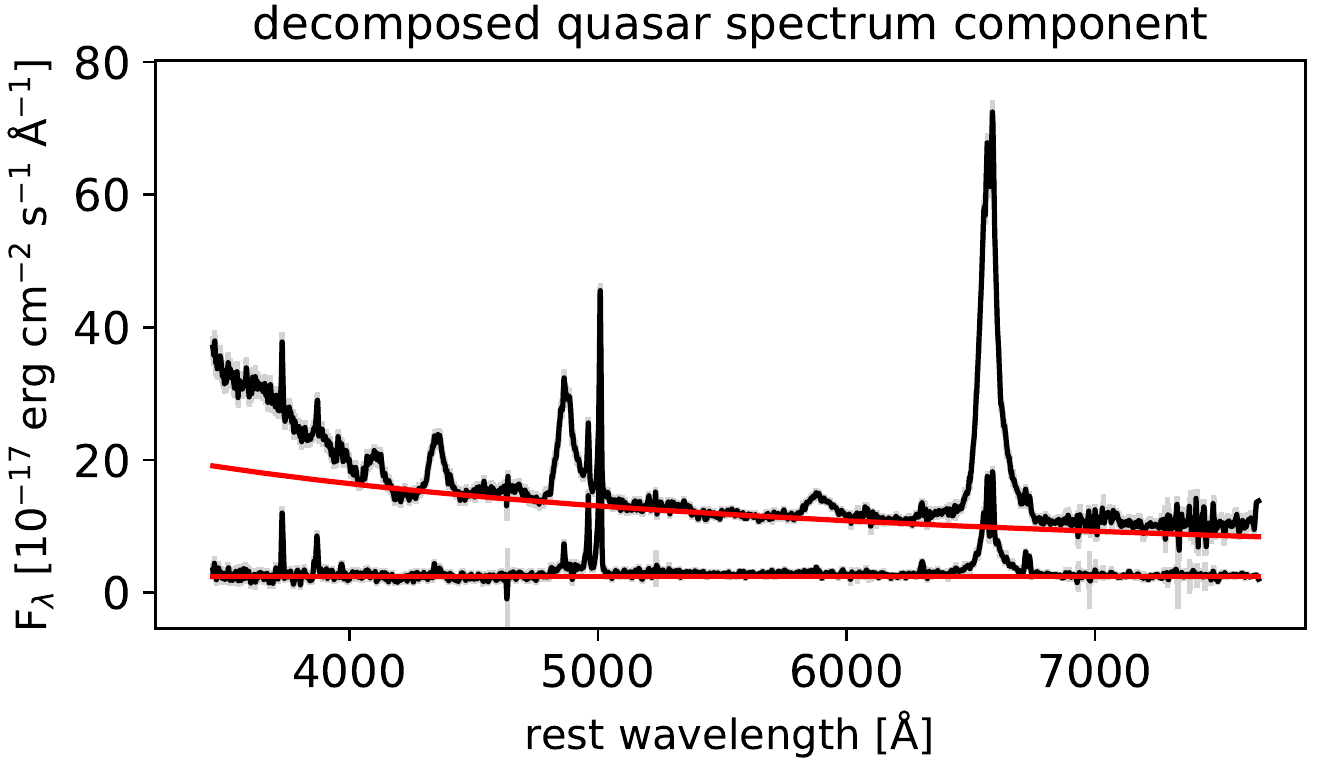}
}
\figcaption{{\bf Spectral decomposition, broad-line fitting, and continuum fitting of SDSS spectra of the changing-look quasar J1021}. 
}
\label{fig:11}
\end{figure*}

%%%  APPENDIX FIG: 12 %%%%%%%%%%%%%%%%%%%%%%%%%%%%%%%%%%%%%%
\begin{figure*}
\center{
\includegraphics*[width=0.47\textwidth]{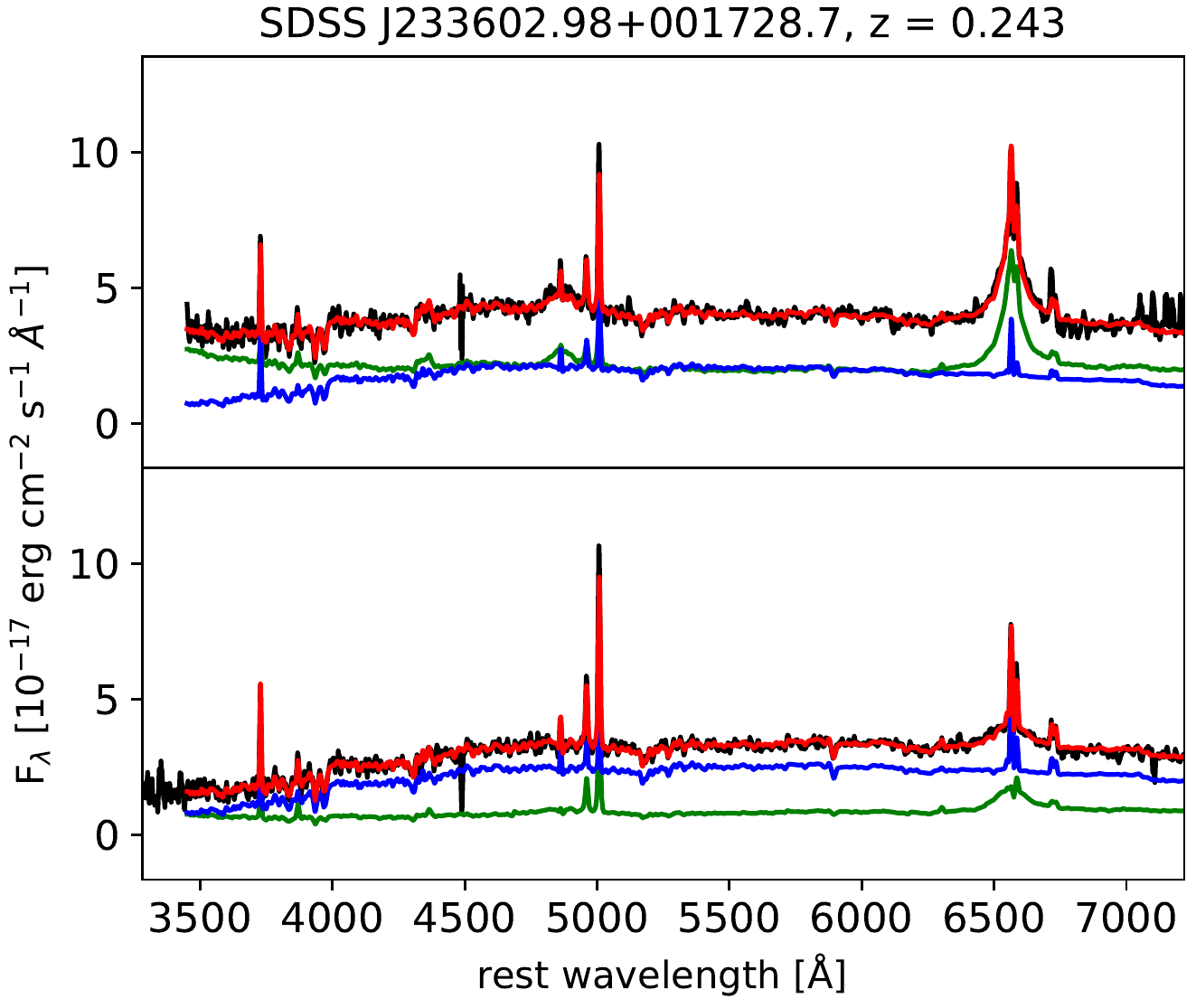} \hspace{10pt}
\includegraphics*[width=0.30\textwidth]{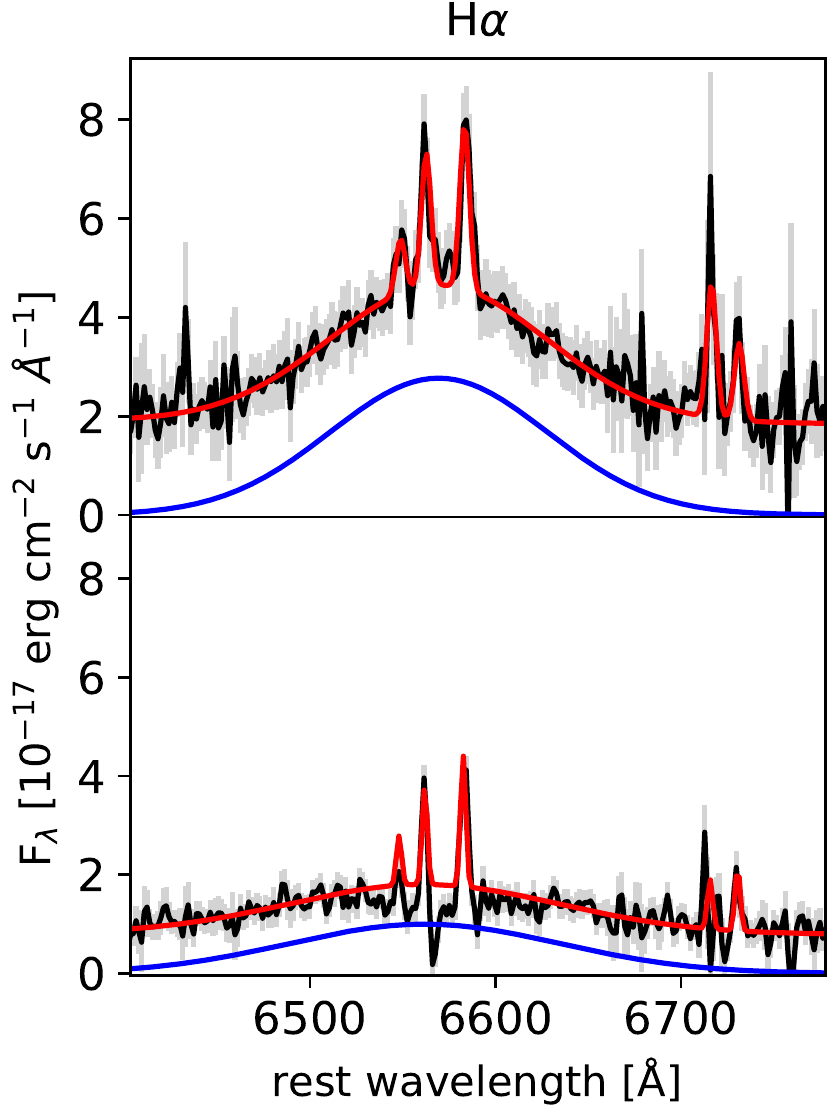}\\ \vspace{10pt}
\includegraphics*[width=0.45\textwidth]{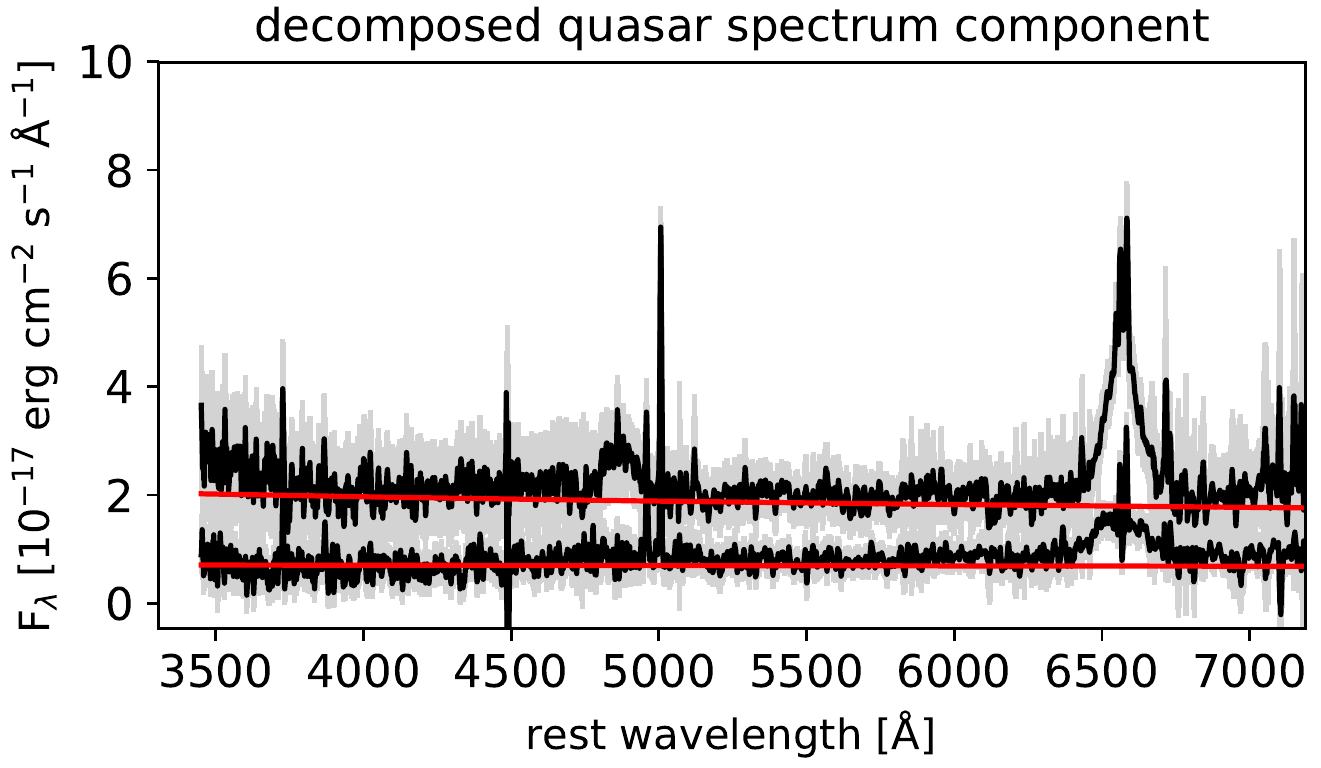}
}
\figcaption{{\bf Spectral decomposition, broad-line fitting, and continuum fitting of SDSS spectra of the changing-look quasar J2336}. 
}
\label{fig:12}
\end{figure*}

\end{appendix}

\end{document}